\documentclass[aps,prd,preprint,groupedaddress]{revtex4}

\usepackage{amsmath}
\usepackage{latexsym}
\usepackage{dsfont}
\usepackage{epsfig}
\usepackage{graphicx}
\usepackage{psfrag}
\usepackage{subfigure}
\usepackage{placeins} 

\newcommand{\unitop}{\mathds{1}}
\newcommand{\bra}[1]{\left\langle #1 \right|}
\newcommand{\ket}[1]{\left| #1 \right\rangle}
\newcommand{\Tr}[1]{\mathrm{Tr} \left[ \, #1 \, \right]}
\newcommand{\Real}[1]{\mathrm{Re}\Big( \, #1 \, \Big)}
\newcommand{\Imag}[1]{\mathrm{Im}\Big( \, #1 \, \Big)}
\newcommand{\bea}{\begin{eqnarray}}
\newcommand{\eea}{\end{eqnarray}}
\newcommand{\be}{\begin{equation}}
\newcommand{\ee}{\end{equation}}
\newcommand{\wh}[1]{\widehat{#1}}
\newcommand{\nn}{\nonumber}
\newcommand{\onehalf}{\frac{1}{2}}
\newcommand{\eqnref}[1]{Eq.~(\ref{#1})}

\newcommand{\figref}[1]{Fig.~\ref{#1}}

\newcommand{\tabref}[1]{Table~\ref{#1}}

\newcommand{\secref}[1]{Sec.~\ref{#1}}

\def\BA{\begin{eqnarray}}
\def\BE{\begin{equation}}
\def\EA{\end{eqnarray}}
\def\EE{\end{equation}}

\def\gtsim{\lower-0.45ex\hbox{$>$}\kern-0.77em\lower0.55ex\hbox{$\sim$}}
\def\ltsim{\lower-0.45ex\hbox{$<$}\kern-0.77em\lower0.55ex\hbox{$\sim$}}

\begin{document}


\title{Formulating Light Cone QCD on the Lattice}


\author{D. Gr\"unewald}
\email[]{d.gruenewald@tphys.uni-heidelberg.de}
\affiliation{Institut f\"ur Theoretische Physik, Universit\"at Heidelberg,
Germany}
\author{E.-M. Ilgenfritz}
\email[]{ilgenfri@physik.hu-berlin.de}
\affiliation{Institut f\"ur Physik, Humboldt-Universit\"at zu
Berlin, Germany}
\author{E.V. Prokhvatilov}
\email[]{Evgeni.Prokhvat@pobox.spbu.ru}
\affiliation{Department of Theoretical Physics, St. Petersburg University,
Russia}
\author{H.J. Pirner}
\email[]{pirner@tphys.uni-heidelberg.de}
\affiliation{Institut f\"ur Theoretische Physik, Universit\"at Heidelberg,
Germany}
\affiliation{Max-Planck-Institut f\"ur Kernphysik Heidelberg, Germany}


\date{\today}

\begin{abstract}
We present the near light cone Hamiltonian $H$ in lattice 
QCD depending on the parameter
$\eta$, which gives the distance to the light cone. 
Since the vacuum has zero momentum 
we can derive an effective Hamiltonian $H_{eff}$ from $H$
which is only quadratic in the momenta and therefore solvable 
by standard methods.
An approximate ground state 
wave functional is determined variationally in the limit $\eta \rightarrow
0$.
\end{abstract}



\maketitle

\section{Introduction}
The lattice approach to QCD pioneered by Wilson~\cite{wilson} and first realized numerically 
by Creutz~\cite{creutz} is based on the QCD action. Moreover, it has been mainly developed in an Euclidean path integral formulation. In contrast to that,
Hamiltonian techniques have remained less studied.
With the Hamiltonian, one can  project out the correct ground state by evolving an initial wave functional 
in imaginary time.
In continuum theory, some progress has been made recently 
in the non-perturbative regime~\cite{Feuchter:2004mk,Leigh:2005dg,
Greensite:2007ij}.
Accoring to these circumstances there have been only few contacts between
lattice QCD and light cone field theory (LCFT).

There is no doubt that LCFT is an important tool for the description of high energy interactions. 
The knowledge of wave functionals in the gauge field configuration space 
may help to calculate light cone wave functions of hadrons.
In the following paper we attempt to take advantage of lattice methods in LCFT (for 
previous work, see~\cite{Mustaki:1988gi,Bardeen:1979xx,Burkardt:2001jg,Dalley:2003aj}). 
Although the Hamiltonian is not Lorentz invariant,
the light cone Hamiltonian \cite{Burkardt:2001jg,pauli} 
offers the advantage of being boost invariant and has -- naively interpreted --  
a trivial vacuum. On the other hand, one  would be surprised if QCD looses its 
non-perturbative vacuum structure in the light cone limit. In our
opinion much of the complicated vacuum structure of QCD is hidden in
the constraint equations appearing in light cone QCD. The constraint equations contain  
zero mode solutions which are difficult to solve. These quantum 
constraint equations have been attacked in lower dimensions for scalar theories, but
gauge theories still escape a solution in higher dimensions. In Nambu Jona Lasinio models 
\cite{Lenz:2004tw} one has been able to
solve these zero mode equations in the large $N_c$ approximation.

A quantization of scalar light cone field theory on the lattice has been first analysed 
in ref. \cite{Mustaki:1988gi} where also the time coordinate has been discretized.
In this reference, special care has been devoted to the constraints which arise on the 
light cone. This approach has not found applications. In particular, it is not easily 
extendable to gauge theories. 

Remarkable progress has been made in light cone QCD with a color dielectric 
lattice theory as a starting point \cite{Bardeen:1979xx,Dalley:2003aj, Dalley:2003uj}. This
approach is based on ``fat" links which arise from averaging gluon
configurations by a block spinning procedure \cite{Mack:1983yi,Pirner:1991im}. 
With this method  the spectrum of glue balls and the pion light cone wave 
function have been calculated \cite{Dalley:2002nj}. In a Lagrangian
framework the connection to the original QCD Lagrangian can be easily made,
although the numerical accuracy is limited. On the light cone, however,
one is prevented from approaching the continuum limit, since an effective 
potential for the link matrices $M\in GL(N)$ with a non vanishing vacuum expectation value 
is not allowed. The norm of the link matrices  $M\in GL(N)$, however, should
approach unity in the continuum limit. 

This is the reason why we propose to formulate QCD near the 
light cone. 
We have already analyzed scalar theories \cite{Prokhvatilov:1994dm}
and QCD \cite{Naus:1997zg,Ilgenfritz:2000bj}  
approaching the light cone in a tilted near light cone 
reference system containing a parameter $\eta \neq 0$ parameterizing the distance 
to the light cone.
Our work in this paper will follow this idea deriving a lattice Hamiltonian which describes  
the pure gauge sector of QCD and which is suitable for a numerical treatment. 
For QCD, we have already followed the path of maximal gauge 
fixing \cite{Naus:1997zg,Ilgenfritz:2000bj} outlined by the Erlangen group \cite{lenz} 
in previous works. This way to 
eliminate all gauge degrees of freedom looks very attractive analytically, but numerically 
it is  not advantageous. It includes solutions of constraint equations which 
complicate the form of the Hamiltonian. Hence we do 
not fix the gauge in the following work and try to establish a form of the Hamiltonian
describing the near light cone dynamics similar to the QCD Hamiltonian in an equal time approach, 
i.e. in terms of unitary matrices describing the gauge degrees of freedom and their 
canonically conjugate momenta.
In our lattice prescription, we leave {\it near light cone} time continuous. 
It plays a similar role as ordinary Minkowski time,
therefore, we can follow the conventional method of the transfer matrix in order to derive the lattice 
Hamiltonian from the lattice action.  
The transversal
field strengths are increased in magnitude due to the boost into the vicinity of the light 
cone whereas the longitudinal fields remain unchanged. Constraint equations arise in the light cone
Hamiltonian framework, since the Lagrangian contains the velocities in
linear form. The momenta related to these velocities obey constraint equations. 
The constraint equations appear in the near light cone
Hamiltonian as terms proportional to $1/\eta^2$.
These terms enforce the ``equality" of the transverse 
 chromo electric and chromo magnetic fields $E_k^a=F_{-k}^a$.
While the longitudinal chromo electric field 
and the longitudinal chromo magnetic field appear in their usual form in the
light cone Hamiltonian, the Hamiltonian contains the transverse chromo magnetic field 
squared in an unusual quadratic $Z(2)$ invariant form. The $Z(2)$ invariance, however, is broken
because the chromo magnetic fields also appear linearly together with the transverse chromo electric field.

The lattice Hamiltonian density depends on an effective constant which
represents the product of the anisotropy parameter $\xi=a_-/a_{\bot}$
and the near light cone parameter $\eta$. If one chooses $\eta=1$ and lets $\xi \rightarrow 0$
 one obtains a deformed system which is squeezed in the spatial ($-$)-direction, if one uses $\xi=1$
and lets $\eta \rightarrow 0$ one obtains the light cone limit. This 
equivalence has been advocated before by Verlinde and Verlinde
\cite{Verlinde:1993te} and Arefeva \cite{Arefeva:1993hi}.
These authors have proposed to implement the strong interaction with
such asymmetric lattices in order to study high energy scattering,
motivating us to proceed in this way.  
As it stands,
the (anisotropic) lattice Hamiltonian itself is not usable for Monte Carlo methods evolving
an arbitrary initial state in imaginary time to the ground state, since the chromo electric
field strengths i.e. the momenta canonically conjugate to the links appear linearly. Therefore we propose to use the
translational invariance of the vacuum to add a term $1/\eta^2 P_-$ in order
to cancel the unwanted terms. Naively this amounts to returning to an
effective lattice 
Hamiltonian which is proportional to the energy in ordinary Minkowski coordinates. 
For the ground state of the vacuum this seems a reasonable procedure. Applications of the
light cone coordinates in finite temperature field theory have followed the same
route \cite{Raufeisen:2004dg}. The new effective Hamiltonian contains two parts:
The first describing the dynamics of the longitudinal chromo electric and chromo magnetic fields 
is not influenced by the smallness of the near light cone parameter $\eta$.
The second part containing the transverse chromo electric and chromo magnetic terms is
enhanced with $\eta$ in the light cone limit.

We analytically investigate this effective Hamiltonian in the strong
and weak coupling limit. Such a  procedure can direct the search for an appropriate 
(approximate) guidance wave functional needed to improve the convergence of the Hamiltonian Monte Carlo method.
The strong coupling limit suggests a simple sum of plaquette terms with different
weights in the purely transverse and ($-$)-transverse planes. The magnitude of the
couplings follows the asymmetries existing in the Hamiltonian. In the limit $\eta \rightarrow 0$
the plaquette terms in the purely transverse planes are weighted very weakly, i.e. the longitudinal
magnetic fields can vary freely. The weak coupling
approximation identifies the ``abelian" fluctuations with their modified dispersion
relations following the built in anisotropy. It is particularly interesting that the light cone
limit $\eta \rightarrow 0$ produces 
long-range correlations in the minus direction, which deviate from the local strong coupling 
ansatz. In fact this anisotropy may help to make an ansatz for the ground state wave functional
which is especially appropriate in the light cone limit. It correlates fluctuations of the 
longitudinal chromo magnetic fields with long strings along the ($-$)-direction. This ansatz may also
point the way to find a solution of the quantum constraint of the initial Hamiltonian. 

The outline of the paper is as follows: In \secref{sec:ContHandP} we introduce near
light cone coordinates. For the sake of clarity, we first establish the 
methodology in the continuum formulation. We derive the continuum Hamiltonian and
momentum operator. Furthermore, we motivate an effective Hamiltonian
making an ordinary Quantum Diffusion Monte Carlo algorithm possible.
In \secref{sec:HamPMinus} we switch to the lattice formulation and derive the near light cone 
Hamiltonian from the latticized action with the transfer matrix method. In \secref{sec:EffNLCH} 
we set up the effective Hamiltonian. The time independent Schr\"odinger equation for the effective Hamiltonian is 
analytically solved for the ground state in the strong and weak coupling limit in \secref{sec:AnalyticalSolutions}. In \secref{sec:VarOpt} we
variationally optimize an ansatz for the ground state wave functional motivated
by the strong and weak coupling analysis. It allows to interpolate between
these two extreme limits and to investigate the $\eta$ behavior in the whole
coupling range.  
Finally, in \secref{sec:OutlookAndConclusions}, we present our conclusions and an outlook to
future work.

\section{The continuum QCD Hamiltonian and momentum near the light cone}
\label{sec:ContHandP}
Before we start with the actual derivation of the QCD Hamiltonian and
momentum near the light cone, we would like to introduce near light
cone coordinates similar to the coordinates first proposed by
\cite{Prokhvatilov:1989eq,Lenz:1991sa}.  The transition to near light
cone (nlc) coordinates might be considered as a two-step process.
In the first step, one starts in ordinary Minkowski space in the laboratory frame
with unprimed coordinates $x^\mu$ and transforms into a reference
frame described by primed coordinates $x^{\prime \mu}$ which moves 
with relative velocity $\beta$ along the longitudinal direction relative
to the laboratory frame. The relative velocity $\beta$ is chosen to be given
by
\bea \beta &=&
\frac{1-\eta^2}{1+\eta^2} \;,\; \eta~\in~\left[0,1\right].  
\eea 
The associated Lorentz transformation expressing the primed 
coordinates in terms of laboratory frame coordinates reads 
\bea 
x^{\prime 0} &=& \gamma\left(x^0-\beta
\, x^3 \right) \nn\\ x^{\prime 3} &=& \gamma\left(x^3-\beta \, x^0
\right) \;,\; \gamma=\left(1-\beta^2\right)^{-1/2} \; .
\eea 
Here $x^0$ and $x^3$ denote
the temporal coordinate and the longitudinal spatial coordinate
respectively in usual Minkowski coordinates.  From the boosted frame,
one performs an additional linear transformation not included in the
Lorentz group which rotates the temporal and longitudinal coordinates. 
It is given by 
\BA
x^+&=&\frac{1}{2}\Bigl[\left(1+\eta^2\right)x^{\prime
0}+ \left(1-\eta^2\right)x^{\prime 3}\Bigr] \nonumber\\
x^-&=&\phantom{\frac{1}{2}}\Bigl[x^{\prime 0}-x^{\prime 3}\Bigr] \; .
\label{eqn:TrafoNLC}
\EA
Here, $x^+$ is defined to be the new time coordinate along which the system evolves and 
$x^-$ is defined to be the new spatial longitudinal coordinate. 
The transversal coordinates $x^1$ and $x^2$ remain unchanged. By quantizing
a theory on a hypersurface of constant $x^+$, one can smoothly interpolate between
an equal time quantization and light cone quantization by varying the external near light cone
parameter $\eta$ from $1$ to $0$.
In the equal time limit $\eta=1$, the temporal coordinate $x^+$ is given by the ordinary Minkowski time 
coordinate $x^+ = x^{0\prime}$ and $\beta=0$, i.e. the new reference frame is not moving relative to the 
laboratory frame. 
In the light cone limit $\eta \rightarrow 0$, 
$x^+$ is proportional to the usual temporal light cone coordinate and $\beta$ approaches 
$\beta=1$.
The nlc energy $p_+$ and longitudinal momentum $p_-$ expressed in terms of the 
laboratory energy $E$ and longitudinal momentum $p^3$ are given by
\bea
p_+&=& \frac{1}{\eta}\left(E-p^3\right)\nn\\ 
p_-&=& \eta \, p^3  \; .                    
\label{eqn:MomentaRel}
\eea 
The second relation in \eqnref{eqn:MomentaRel} shows that the
magnitude of longitudinal momenta is reduced by transforming to nlc
coordinates. 
In other words, large longitudinal momenta in the lab frame $p^3 \propto 1 /(a_- \eta)$
become accessible by a nlc lattice with longitudinal lattice spacing $a_-$ for $\eta \rightarrow 0$. This makes nlc coordinates physically very attractive.
  
The definition of 
nlc coordinates \eqnref{eqn:TrafoNLC} induces the following metric:
\begin{equation}
 \begin{array}{cc}
  g_{\mu\nu}=\left(
   \begin{array}{cccc}
      0 &  0 &  0 &  1 \\
      0 & -1 &  0 &  0 \\
      0 &  0 & -1 &  0 \\
      1 &  0 &  0 &  -\eta^2 
   \end{array}
             \right)  &
  g^{\mu\nu}=\left(
   \begin{array}{cccc}
     \eta^2 &  0 &  0 &  1 \\
          0 & -1 &  0 &  0 \\
          0 &  0 & -1 &  0 \\
          1 &  0 &  0 &  0
   \end{array}
             \right) 
   \end{array}
\label{eq.(2)}   
\end{equation}
with $\mu,\nu=+,1,2,-,\det g=1$. This defines the scalar product
\BA
x_\mu y^\mu & = & x^-y^+ + x^+y^- - \eta^2 x^-y^- - \vec x_\perp\vec
y_\perp\nonumber\\ 
            & = & x_-y_+ + x_+y_- + \eta^2 x_+y_+ - \vec x_\perp\vec
y_\perp \; .
\label{eq.(3)}
\EA
Note, that the metric has off-diagonal terms which implies that there are
terms mixing temporal and longitudinal spatial coordinates in the scalar product.
This has severe consequences for a standard Euclidean lattice approach. 

If we put a pair of color charges propagating along the longitudinal coordinate
$x^-$ described by a longitudinally extended Wegner-Wilson loop and a stationary target 
modeled by a transversal plaquette at fixed $x^-$ in this reference frame, we can simulate 
color dipoles colliding with a hadron \cite{Iancu:2002xk} in the light cone limit. 
In the described way one might be able to calculate cross sections between hadrons.  
For $\eta \rightarrow 0$, we approach the light cone from space like distances which 
is different from the approach of Balitsky \cite{Babansky:2002my} who approaches the light cone
from time like distances closer to scattering experiments. 

For QCD, the pure gluonic part of the Lagrange density in manifestly covariant notation is given by
\bea
\mathcal{L}
           &=&-\frac{1}{4} F_{\mu\nu}^a g^{\mu\kappa}g^{\nu\rho}F_{\kappa\rho}^a \; ,
\label{eqn:LagrangeDens}
\eea
with the non-abelian field strength tensor 
\bea
F_{\mu\nu}^a &\equiv& \partial_{\mu} A_{\nu}^a -
                      \partial_{\nu} A_{\mu}^a + 
                      g f^{abc} A_{\mu}^b A_{\nu}^c  \; .
\eea

In the following, we restrict ourselves to the color gauge group
$SU(2)$ for which the structure constants $f^{abc}$ are given by the
three-dimensional totally antisymmetric Levi-Cevita symbol
$\epsilon^{abc}$.  By using the nlc metric \eqnref{eq.(2)} we obtain
for the Lagrange density \eqnref{eqn:LagrangeDens} \bea \mathcal{L}&=&
\sum\limits_a \left[ \onehalf F_{+-}^aF_{+-}^a
  +\sum\limits_{k=1}^2\left(F_{+k}^aF_{-k}^a+
    \frac{\eta^2}{2}F_{+k}^aF_{+k}^a \right) -\onehalf
  F_{12}^aF_{12}^a \right] \; .
\label{eqn:LagrangeDensnlc}                     
\eea 
Note, that there is a term in the Lagrange density which is only
linear in one of the temporal field strengths, namely
$F_{+k}^aF_{-k}^a$. Therefore, the numerical standard approach for
lattice gauge theory, the Monte Carlo sampling of the Euclidean path
integral does not apply for nlc coordinates. The reasoning is as
follows.  If one performs in analogy to equal time theories an
analytical continuation to imaginary nlc time $x^+\rightarrow -
\mathrm{i} x^+_E$, each temporal field strength is replaced by its
Euclidean counterpart times an additional factor $\mathrm{i}$.
Therefore, the linear term yields a complex valued Euclidean action
and the integrand of the Euclidean path integral is no longer
interpretable as a probability density.  A similar problem arises for
lattice gauge theory at finite baryonic densities which is usually
referred to as the sign problem. So far, no convenient solution has
been found.  In order to avoid these problems, we stay in Minkowski
time for the rest of the paper and we switch to a Hamiltonian
formulation. We perform a Legendre transformation to switch to a Hamiltonian formulation
, i.e. we have to express the
temporal derivatives of the fields by their canonical conjugate
momenta in particular which are given by the functional derivatives of the Lagrange
density with respect to the temporal derivative of the correspondent
fields: 
\bea 
\Pi_{\mu}^a &\equiv& \frac{\delta \mathcal{L}}{\delta
  \left( \partial_+ A_{\mu}^a \right)} \; .  
\eea 
Therefore, the
canonical momenta conjugate to the gauge fields are given by 
\bea
\Pi_k^a &=& \frac{\delta \mathcal{L}}{\delta\partial_+ A_k^a} =
\frac{\delta \mathcal{L}}{\delta F_{+k}^a}
= F_{-k}^a+\eta^2 F_{+k}^a \; , \nn \\
\Pi_-^a &=& \frac{\delta \mathcal{L}}{\delta\partial_+ A_-^a} =
\frac{\delta \mathcal{L}}{\delta F_{+-}^a} = F_{+-}^a \; .
\label{eqn:CanMom}                   
\eea 
Here, we have chosen the axial gauge $A_+^a=0$ which is quite
natural because the temporal gauge field $A_+^a$ is not dynamical,
i.e. there is no temporal derivative appearing in the Lagrange
function.  It acts like a Lagrange multiplier which multiplies the
Gauss law $G=0$ with $G$ given by: 
\bea 
G&=&D_-^{ac} \; \Pi_-^c +D_k^{ac} \; \Pi_k^c\;. 
\eea 
Here $D_\mu^{ac}$ denotes the ordinary
covariant derivative - in the adjoint representation - in spatial
direction $\mu$ 
\bea D_\mu^{ac}&=& \partial_\mu \delta^{ac} + g
f^{abc} A_\mu^b \;.
\eea 
In order to recover the full Lagrangian dynamics,
we have to supplement the equations of motion by Gauss' law. Hence,
the Gauss law has to be imposed as a constraint equation on physical
states. We express the temporal derivatives of the gauge fields in
terms of the canonical conjugate momenta by using \eqnref{eqn:CanMom},
which yields 
\bea
\partial_+ A_k^a &=& F_{+k}^a = 
\frac{1}{\eta^2} \left(\Pi_k^a - F_{-k}^a\right) \; , \nn\\
\partial_+ A_-^a &=& F_{+-}^a = \Pi_-^a \; .
\label{eqn:TDerByMom}
\eea
We may obtain the QCD Hamiltonian and the momentum operator via the energy momentum tensor,
where we have to substitute the temporal derivatives of the gauge fields by the 
corresponding expressions involving the canonical conjugate momenta \eqnref{eqn:TDerByMom}.
If the Lagrange density for an arbitrary field theory with fields $\Phi_r$ 
defined by the Lagrangian density $\mathcal{L}$ is a function of the fields itself 
and derivatives of the fields only, namely $\mathcal{L}=\mathcal{L}(\Phi_r,\partial_{\mu}\Phi_r)$,
the energy momentum tensor in its most general form is given by
\bea
T^{\mu\nu}&=&\sum\limits_r 
\frac{\delta \mathcal{L}}{\delta\left(\partial_{\mu} \Phi_r\right)} 
\partial^{\nu} \Phi_r - g^{\mu\nu} \mathcal{L} \; ,
\eea
It defines the Hamiltonian density $\mathcal{H}$ and the longitudinal momentum density 
$\mathcal{P}_-$ by
\bea 
\mathcal{H}  &=&T_{\;\;\;+}^+ \; , \nn\\
\mathcal{P}_-&=&T_{\;\;\;-}^+ \; .
\eea 
Therefore, for the nlc QCD Lagrangian \eqnref{eqn:LagrangeDensnlc} we find the Hamiltonian density
\bea 
\mathcal{H} &=& \onehalf \sum\limits_a 
\left[ \Pi_-^{a}\Pi_-^{a}+F_{12}^aF_{12}^a
+ \sum\limits_{k=1}^2 \frac{1}{\eta^2}\left(\Pi_k^{a}-F_{-k}^a\right)^2
                                       \right]
\label{eqn:HamiltonianDensity}
\eea
and the longitudinal momentum density
\bea
\mathcal{P}_-&=& \Pi_-^a \partial_- A_-^a +
\sum\limits_{k=1}^2 \Pi_k^a \partial_- A_k^a  \; .
\eea 
This form of the local integrand for the generator $\mathcal{P}_-$ of longitudinal translations 
is not manifestly gauge invariant. 
However, if one uses Gauss' law and the definition of the field 
\-strength tensor one can rewrite $\mathcal{P}_-$ as
\bea
\mathcal{P}_-&=& \Pi_k^aF_{-k}^a+ \partial_k \left(\Pi_k^aA_{-}^a \right)
                                + \partial_- \left(\Pi_-^aA_{-}^a \right) \; .
\eea 
So, the longitudinal momentum density $\mathcal{P}_-$ 
may be expressed as a manifestly gauge invariant object plus 
some total derivatives along the spatial directions which disappear
after integration with periodic boundary conditions. We use the symmetrized form
\bea
\mathcal{P}_-&=& \onehalf\left(\Pi_k^a F_{-k}^a+F_{-k}^a\Pi_k^a\right) \; .
\eea
The integrated Hamiltonian density ${H}$ is the generator of nlc ``time" 
translations and the integrated longitudinal momentum operator 
${P}_-$ is the generator of spatial translations in longitudinal direction:
\bea
{H}   &=& \int d^2x_{\bot} dx^- {\mathcal{H}} \; , \nn\\
{P}_- &=& \int d^2x_{\bot} dx^- {\mathcal{P}}_- \; .
\label{eqn:HamMom}
\eea 
We quantize the theory by choosing the following commutation relations
at equal light cone time $x^+$ :
\bea
\left[{\Pi}_m^a(\vec{x}), {A}_n^b(\vec{y})\right]&=& 
   -\mathrm{i} \delta^{ab}\delta_{mn}\delta^{(3)}(\vec{x}-\vec{y}) \; , \nn\\
\left[{\Pi}_m^a(\vec{x}), {\Pi}_n^b(\vec{y})\right]&=& 0 \; , \nn\\   
\left[{A}_m^a(\vec{x}), {A}_n^b(\vec{y})\right]&=& 0 \; .
\label{eqn:ContComm}
\eea These commutator relations respect the Heisenberg equations of
motion.  Note, analogously for Quantum Mechanics, one has to
supplement the Heisenberg equations of motion by the Gauss law
constraint.  In quantum mechanics, the Gauss law constraint translates
into a restriction of the Hilbert space to the subspace of physical
states i.e. states $\Psi$ satisfying the Gauss law 
\bea
\left({D}_-^{ab}{\Pi}_-^b(\vec{x})+\sum\limits_{k=1}^2
  {D}_k^{ab}{\Pi}_k^b(\vec{x})\right) \ket{\Psi}&=&0
\;\;\;\;\forall\;\vec{x},a \; .  
\eea 
Since the Gauss law operator is
the generator of gauge transformations the physical subspace is given
by that part of the entire Hilbert space which is spanned by gauge
invariant states.  \newline The $1/\eta^2 $-term in the Hamiltonian
\eqnref{eqn:HamiltonianDensity} favors expectation values of
transverse chromo electric fields $\Pi_k^{a}$ and transverse
chromo magnetic fields $F_{-k}^{a}$ to be equal in order to have a
minimal energy.  On the other hand, this term introduces terms linear
in momentum which are difficult to handle, 
for example with a numerical Quantum Diffusion Monte Carlo algorithm 
\cite{Chin:1983pc,Chin:1985ua,Heys:1984kg} which exploits the fact that 
the time evolution operator is a projector onto
the ground state when analytically continued to imaginary times. 
The terms linear in the momentum enforce the wave functional to be complex
valued in general which spoils the whole procedure. 
These are exactly the same terms which make the nlc action complex valued
after the Wick rotation in the action based formulation.  
Hence, the problem reappears in the Hamiltonian formulation. 
However, for Hamiltonian nlc 
QCD it is possible to define an effective Hamiltonian converging to
the exact ground state which avoids the problematic terms.  \newline
Obviously, the Hamilton operator ${H}$ in \eqnref{eqn:HamMom} is
translation invariant and gauge invariant. Hence, it commutes with the
longitudinal momentum operator ${P_-}$ and with the Gauss operator
$G$:
\bea
\left[{H},{P}_-\right] &=& 0  \; , \nn \\
\left[{H}, G \right] &=& 0 \; .  
\eea 
Therefore, common eigenstates
exist which diagonalize the Hamiltonian and the longitudinal momentum
operator simultaneously and in addition fulfill the Gauss law. In
particular, momentum is a good quantum number which is left invariant
by time evolution.  In order to solve the Hamiltonian we are
interested in translation-invariant states which are eigenstates of
the longitudinal momentum operator, i.e.  with eigenvalue equal zero.
In vacuum, with light cone momentum $P_-=0$, we can add
$(1/\eta^2)~P_-$ to define an effective Hamiltonian density
$\mathcal{H}_{eff}$ which is only quadratic in momenta: \bea
\mathcal{H}_{eff}&=& \mathcal{H}+\frac{1}{\eta^2}\mathcal{P}_- \nn \\
&=& \frac{1}{2} \sum\limits_a \left[
  \Pi_-^{a}\Pi_-^{a}+F_{12}^aF_{12}^a + \sum\limits_{k=1}^2
  \frac{1}{\eta^2}\left(\Pi_k^{a}\Pi_k^{a} + F_{-k}^aF_{-k}^a\right)
\right] \; .
\label{eq:quadratic}
\eea 
This effective Hamiltonian density is still symmetric under the exchange
\bea
\Pi_k^a \longleftrightarrow F_{-k}^a  \; ,
\eea
but it does not enforce the equality between transverse chromo electric and transverse
chromo magnetic fields commonly used in the light cone limit also for the
quantum field theoretic system.
\newline
One finds the ground state $\ket{\Psi_0}$ of ${H}$ by evolving a translation invariant 
trial state $\ket{\Phi}$ not orthogonal to $\ket{\Psi_0}$ with the effective 
time evolution operator 
\bea
\ket{\Psi_0} &=&\lim_{\tau \rightarrow \infty} 
             \exp\left[-\left({H}_{eff}-E_{eff}\right)\tau\right] \ket{\Phi}
\label{eqn:EffTimeEvo}
\eea
related to the effective Hamiltonian. For the details of an explicit implementation
of the ground state projection operator with a guided Quantum Diffusion Monte Carlo 
algorithm, we refer the reader to \cite{Chin:1983pc,Chin:1985ua,Heys:1984kg}. 
In order to direct the Monte Carlo into regions of the con\-figuration space which have 
large acceptance rates, i.e. which have a large exact ground state probability density
for the given configuration, one introduces a guidance wave functional which is an approximation 
of the exact ground state. Instead of evolving the trial state itself, one evolves a 
probability density in imaginary time which converges to the product of the exact ground 
state wave functional and the guidance wave functional for asymptotic times. Obviously, 
the application of a guidance wave functional introduces some bias in the computation of expectation 
values. However, in principle one can get rid of this bias by applying forward walking techniques 
\cite{Ceperley:1979,Hamer:2000em}. 
\newline
The algorithm preserves Gauss' law as long as the guidance/trial wave functional is a functional of
closed loops only. In principle, there are also 
multiple connected loops possible, but then the
chromo electric flux must be conserved at each site. In this paper
our primary objective is to translate the discussed methodology onto
the lattice and to determine variationally a 
good starting and guidance wave functional $\ket{\Phi}$ for the Quantum Diffusion Monte 
Carlo evolution based on \eqnref{eqn:EffTimeEvo} which is motivated by analytical computations in the 
strong and weak coupling limit.

\section {Near light cone Hamiltonian ${H}$ on the lattice}
\label{sec:HamPMinus}
In order to regularize the continuum formulation we go over to the lattice formulation.
In a previous paper \cite{Ilgenfritz:2006ir} we have started deriving  
the Hamiltonian for gauge theories on the lattice from the path integral
via the transition $a_+\rightarrow 0$. 
Here, we go through the procedure in detail. We introduce in four-dimensional
space $SU(N)$ link variables $U_i(x)$ connecting the site $x$  
with the 4D site $x+\wh{e}_i$ ($\wh{e}_i$, $i=+,-,1,2$, is a unit vector 
in 4D space-time) in the following way:
\bea
U_i(x)&\equiv& \mathcal{P}\exp\left(\mathrm{i}~g \int\limits_{x}^{x+\wh{e}_i}
   dy^{\mu}~A_{\mu}^a(y)~\frac{\sigma_a}{2} \right) \; ,
\label{eqn:PrimDefLink}  
\eea
where $\mathcal{P}$ is implementing path ordering 
from left to right with increasing $y^\mu$  
and $\sigma_a$ are hermitian color generators. In the following, we
restrict to $SU(2)$ where the $\sigma_a$ are given by the Pauli matrices.
The hermitian conjugate of the link variables, $U_i^{\dagger}(x)$, connect 
the site $x+\wh{e}_i$ with the site $x$ in reverse order.
Plaquettes $U_{ij}(x)$ are related to the field strengths $F_{ij}(x)$ and have the usual form
\bea
U_{ij}(x)=U_{i}(x)U_{j}(x+\wh{e}_i)U_{i}^{\dagger}(x+\wh{e}_j)U_{j}^{\dagger}(x)~.
\eea
Expanding a plaquette around its center $x+\frac{\wh{e}_i}{2}+\frac{\wh{e}_j}{2}$ 
in orders of the 
lattice spacing one obtains 
\bea
U_{ij}(x)=
\unitop+\mathrm{i} g a_i a_j F_{ij}^a \frac{\sigma_a}{2} -
\onehalf g^2 a_i^2 a_j^2 F_{ij}^a F_{ij}^b \frac{\sigma_a}{2} \frac{\sigma_b}{2} +\ldots~.
\eea
Here $a_i$ denotes the lattice spacing for direction $i$, i.e. we allow in general for different 
lattice spacings in the temporal, longitudinal and transversal directions. 
A correspondent expansion is obtained for $U_{ij}^{\dagger}(x)$.
Therefore, the sum over color indices $a=b$ of a product of field strengths  
is given in the limit $a_i \rightarrow 0 $ as follows
\bea
F_{ij}^a F_{ij}^a(x) = \frac{4}{g^2 a_i^2 a_j^2}~\Tr{ \unitop - \Real{U_{ij}(x)}} \; .
\label{eqn:LatticeFijsquared}  
\eea
Or more general 
\bea
F_{ij}^a F_{kl}^a(x) = \frac{2}{g^2 a_i a_j a_k a_l} 
                   \Tr{\Imag{ U_{ij}(x)}~\Imag{U_{kl}(x)} } \; .
\label{eqn:LatticeFijFkl}
\eea
Here, $\mathrm{Re}\left(U\right) $ and $\mathrm{Im}\left(U\right)$ are defined as   
\bea
\centering
\mathrm{Re}\left(U\right)\equiv\frac{U+U^{\dagger}}{2} & , & 
  \mathrm{Im}\left(U\right)\equiv\frac{U-U^{\dagger}}{2\mathrm{i}} \; .
\eea
In \eqnref{eqn:LatticeFijFkl} the two plaquettes $U_{ij}$ and $U_{kl}$ begin and end at the 
common site $x$.
Note, that \eqnref{eqn:LatticeFijsquared} and \eqnref{eqn:LatticeFijFkl} are
representations of the field strengths squared terms which are only valid in leading order 
of the lattice spacing. So far, there is no improvement included.   
By using the relations \eqnref{eqn:LatticeFijsquared} and \eqnref{eqn:LatticeFijFkl},
we may rewrite the continuum nlc Lagrange density \eqnref{eqn:LagrangeDensnlc} in terms of plaquettes 
such that it is recovered in the naive continuum limit.
Similar to the equal time case \cite{Creutz:1976ch}, one can fix
inside the path integral on the lattice a maximal tree of links to
arbitrary group elements. A maximal tree of links is a tree to
which no more links can be added without forming a loop. 
By doing so, the path integral itself and expectation values of
gauge invariant operators are not affected. Hence, we fix all time-like
links $U_+(x)$ to $U_+(x) \equiv 1$ in the following.  This
corresponds to the temporal gauge $A_+=0$ and one obtains for the lattice 
analog $S_{lat}$ of the action $S=\int d^4x~\mathcal{L}$: 
\bea 
S_{lat} & = & \frac{2}{g^2}
\sum\limits_x \left\{ \frac{a_{\bot}^2}{a_+ a_-}~\Tr{\unitop - \Real{
U_{-}(x+\wh{e}_+)U_{-}^{\dagger}(x) }} \right.  \nn\\ & &
\left. -\frac{a_+a_-}{a_{\bot}^2}~\Tr{\unitop - \Real{U_{12}(x)}}
\right. \nn\\ & & \left. +\sum\limits_{k}~\Tr{\Imag{
U_{k}(x+\wh{e}_+)U_k^{\dagger}(x) }~\Imag{ U_{-k}(x) }} \right. \\ &
&\left.+\frac{a_-}{a_+}~\eta^2\sum\limits_{k}~\Tr{\unitop -
\Real{U_{k}(x+\wh{e}_+)U_k^{\dagger}(x) }} \right\} \, . \nn
\label{eqn:LatticeAction}      
\eea
Therefore, the QCD path integral on the lattice in the $A_+=0$ gauge and with the $SU(2)$ Haar measure $dU$ is given by
\bea
Z&=&\int \left[\prod\limits_{x}\prod\limits_{j=1,2,-} dU_{j}(x)\right]~e^{\mathrm{i} S_{lat}}~.
\eea
In order to obtain the
lattice Hamiltonian, we would like to go over from the action-based
path-integral to a Hilbert-space formulation of the near light cone
QCD lattice gauge theory in the following, letting the time-like
lattice constant approach zero.  The method is similar to the
transition from the action to the Hamiltonian
in ordinary Euclidean $SU(2)$ lattice gauge theory 
carried out by Creutz \cite{Creutz:1976ch}.
 
The procedure consists of two steps. 
First, we construct the transfer matrix $T$.   
Second, we define the space on which it acts and rewrite the transfer matrix in terms of the conjugated momenta of the links and extract the lattice Hamiltonian by identifying the transfer matrix with the time evolution operator which propagates the system from one time slice to the next.

Note, that the lattice action \eqnref{eqn:LatticeAction} is local in
the temporal direction.  Each piece is connecting two adjacent
time slices $x^{\prime+}=x^++a_+$ and $x^+$ which means that the path
integral factorizes into a product of transfer matrices
$T(x^{\prime+},x^+)$.  
\bea 
T & = & \Bigg[\prod\limits_{\vec{x}}
~\exp\left\{\mathrm{i}\frac{2}{g^2}~\frac{a_{\bot}^2}{a_+a_-}~\Tr{
\unitop -
\Real{U_{-}(\vec{x},x^{\prime+})U_{-}^{\dagger}(\vec{x},x^+)}}\right\}\Bigg]
\times \nn \\ & & \Bigg[ \prod\limits_{\vec{x},k} ~\exp\left\{
\mathrm{i}~\frac{2}{g^2}~\eta^2~\frac{a_-}{a_+}~\Tr{ \unitop -
\Real{U_{k}(\vec{x},x^{\prime+})U_{k}^{\dagger}(\vec{x},x^+)}}\right\}\times
\Bigg.  \nn \\ & & \Bigg.  \hspace{.2cm} ~\exp\left\{
\mathrm{i}~\frac{2}{g^2}~\Tr{
\Imag{U_{k}(\vec{x},x^{\prime+})U_{k}^{\dagger}(\vec{x},x^+)}~\Imag{{U}_{-k}(\vec{x},x^+)}}
\right\} \Bigg] \times \nn\\ & & \Bigg[\prod\limits_{\vec{x}}
\exp\left\{-\mathrm{i}~\frac{2}{g^2}~\frac{a_+a_-}{a_{\bot}^2}~\Tr{
\unitop - \Real{{U}_{12}(\vec{x},x^+)}}\right\} \Bigg] \; .
\label{eqn:Top1}
\eea
Here, $\vec{x}$ denotes a lattice vector in the three dimensional spatial sub lattice.
If we denote by the set of links $\mathcal{U}(x^+)$   
an entire spatial lattice configuration at time $x^+$,
the transfer matrix $T$ evolves the 
configuration $\mathcal{U}(x^+)$ 
at the time slice $x^+$ to the 
configuration $\mathcal{U}(x^{\prime+})$ 
at the neighboring time slice $x^{\prime+}$ in our convention. The construction of the Hilbert space and the transcription of the temporal plaquettes in terms of momenta canonically conjugate 
to the links is similar to the steps performed in \cite{Creutz:1976ch}. The interested reader
may find the explicit calculation in appendix~\ref{sec:TransferMatrix}. 
One finally obtains for the lattice Hamiltonian
\bea 
{H}_{\mathrm{lat}}&=& \sum\limits_{\vec{x}} \Bigg[\hspace{-0.14cm}\Bigg[
~\frac{g^2}{2}~\frac{1}{a_-}~\sum\limits_{k,a}~\frac{1}{\eta^2}~
\left\{~{\Pi}_k^a(\vec{x})-\frac{2}{g^2}~\Tr{\frac{\sigma_a}{2}~\Imag{{U}_{-k}(\vec{x})}}
\right\}^2 \Bigg. \Bigg. \nn\\ & & \hspace{0.5cm} \Bigg. \Bigg.
+~\frac{g^2}{2}~\frac{a_-}{a_{\bot}^2}~\sum\limits_{a}{\Pi}_-^a(\vec{x})^2
+~\frac{2}{g^2}~\frac{a_-}{a_{\bot}^2}~\Tr{\unitop-\Real{{U}_{12}(\vec{x})}}
\Bigg]\hspace{-0.14cm}\Bigg] \; . \nn\\ & & 
\eea
Here, the operators $\Pi_k^a(\vec{x})$ are canonically conjugate to the link operators
and they obey the following commutation relations
\bea
\left[ {\Pi}_j^a(\vec{x}),{U}_{j^\prime}(\vec{x}^\prime) \right]
&=& \phantom{-}
\frac{\sigma_a}{2}~{U}_j(\vec{x})~\delta_{j,j^\prime}~\delta_{\vec{x},\vec{x}^\prime}
\; , \nn\\ 
\left[
{\Pi}_j^a(\vec{x}),{U}_{j^\prime}^{\dagger}(\vec{x}^\prime) \right]
&=&
-{U}_j^{\dagger}(\vec{x})~\frac{\sigma_a}{2}~\delta_{j,j^\prime}~\delta_{\vec{x},\vec{x}^\prime}\; ,
\nn\\
\left[ {\Pi}_j^a(\vec{x}),{\Pi}_{j^\prime}^b(\vec{x}^\prime) \right] 
&=&
\mathrm{i}~\varepsilon^{abc}~{\Pi}_j^c(\vec{x})~\delta_{j,j^\prime}~\delta_{\vec{x},\vec{x}^\prime}
\; ,\nn\\ \left[ {\Pi}_j(\vec{x})^2,{\Pi}_{j^\prime}^b(\vec{x}^\prime)
\right] &=& 0\;.
\eea
In analogy to the continuum Hamiltonian density cf. \eqnref{eqn:HamiltonianDensity} we introduce the lattice
Hamiltonian density
\bea
\mathcal{H}_{\mathrm{lat}}&\equiv&\frac{H_{\mathrm{lat}}}{V_{\mathrm{lat}}}
   = \frac{1}{\xi a_{\bot}^3} \frac{H_{\mathrm{lat}}}{N_-N_{\bot}^2} \; .
\eea
The lattice anisotropy parameter $\xi$ is given by the ratio of the longitudinal lattice
spacing to the transversal lattice spacing
\bea
\xi&\equiv&\frac{a_-}{a_{\bot}} \; .
\eea
Furthermore, in order to simplify the notation, we have introduced the coupling constant $\lambda$
which is related to the ordinary $SU(2)$ lattice gauge theory coupling $\beta$ by
\bea
\lambda &\equiv& \frac{4}{g^4}=\left( \frac{1}{2} \beta \right)^2 ~~,~~\beta=\frac{4}{g^2}\;.
\eea
Therefore, we obtain for the Hamiltonian density on the lattice
\bea
\mathcal{H}_{\mathrm{lat}}&=& \frac{1}{N_-N_{\bot}^2}\frac{1}{a_{\bot}^4} \frac{2}{\sqrt{\lambda}}
\sum\limits_{\vec{x}}\left\{
    \sum\limits_{a}~\frac{1}{2}~{\Pi}_-^a(\vec{x})^2~
    +\onehalf~\lambda~\Tr{\unitop-\Real{{U}_{12}(\vec{x})}}
    \right.  \nn\\
& &  \left.
    ~+\sum\limits_{k,a}~\frac{1}{2}~\frac{1}{\xi^2\eta^2}~ 
      \Bigg[~{\Pi}_k^a(\vec{x})-\sqrt{\lambda}~\Tr{\frac{\sigma_a}{2}~\Imag{U_{-k}(\vec{x})}}
      \Bigg]^2 
    \right\}\; . 
\label{eqn:LatticeHamiltonianDensity}
\eea 
One observes that the energy density $\mathcal{H}_{\mathrm{lat}}$
only depends on the effective constant $\tilde{\eta}$ defined as the product of the
anisotropy parameter $\xi=a_-/a_\bot$ with $\eta$ instead of both of
them separately
\bea
\tilde{\eta} &\equiv& \xi\cdot\eta \;.
\eea 
Very clearly one can vary two independent parameters $\lambda=4/g^4$ and $\tilde{\eta}$.
The $\tilde{\eta}$ variation may be interpreted in two parametrically distinct but physically 
equivalent ways. 
If one chooses $\eta=1$ and varies $\xi$, one simulates an effective equal time theory
with a ratio of lattice constants $\xi=a_-/a_{\bot}$. In the limit $\xi \rightarrow 0$ 
one ends up with a system,
which is contracted in the longitudinal direction. Verlinde and
Verlinde \cite{Verlinde:1993te} and Arefeva \cite{Arefeva:1993hi} 
have advocated such a set-up to describe high
energy scattering. A contracted longitudinal system means that the minimal 
momenta become high in longitudinal direction and this looks a promising
starting point for high energy scattering. It is obvious that this limit
leads to the same physics as
the limit $\eta \rightarrow 0$ and $\xi=1$, i.e. as the light cone 
limit with equal lattice constants in longitudinal and transverse directions.
\newline
In both limiting cases, i.e. for $\tilde{\eta} \rightarrow 0$, 
the near light cone Hamiltonian is dominated by the term proportional 
to $(1/\tilde{\eta}^2)$. 
Therefore, in the light cone limit the transverse chromo electric fields $\Pi_k$ should 
become equal to the scaled transverse chromo magnetic fields 
$\mathrm{Tr}[\sigma^a/2~\mathrm{Im}(U_{-k})]$. This is a form of electro-magnetic
duality characteristic of light cone gauge field theory. 
\newline
In the following we set $\xi=1$ bearing
in mind that the physical ratio of longitudinal to transverse
lattice spacings for $\eta\neq 0$ may be modified by quantum corrections from
the QCD dynamics.

\section{ Effective near light cone lattice Hamiltonian}
\label{sec:EffNLCH}
To obtain the same cancellation of linear terms in $\Pi_k$ in the effective lattice 
Hamiltonian as in
the continuum \eqnref{eq:quadratic} in order to make a guided Diffusion Monte Carlo in principle possible, 
we add $\mathcal{P}_{-,\mathrm{lat}}$ to the lattice Hamiltonian density
\bea
\mathcal{H}_{\mathrm{eff,lat}}=\mathcal{H}_{\mathrm{lat}}+\frac{1}{\eta^2}\mathcal{P}_{-,\mathrm{lat}}\; . 
\eea
Here the density $\mathcal{P}_{-,\mathrm{lat}}$ is defined as
\bea
\mathcal{P}_{-,\mathrm{lat}}&\equiv&\frac{P_{-,\mathrm{lat}}}{V_{\mathrm{lat}}} \nn\\
&=&\frac{1}{N_-N_{\bot}^2} 
\frac{1}{\xi^2}\frac{1}{a_{\bot}^4} 
\sum\limits_{\vec{x},k,a} \left( \Pi_k^a(\vec{x}) \cdot \Tr{\frac{\sigma_a}{2}~\Imag{{U}_{-k}(\vec{x})}}
                                        \right. \nn\\
    & &                           
\left.\phantom{\Pi_k^a(\vec{x})\Pi_k^a(\vec{x})\Pi_k^a(\vec{x})}                                
+\Tr{\frac{\sigma_a}{2}~\Imag{{U}_{-k}(\vec{x})}} \cdot \Pi_k^a(\vec{x})\right)\; .
\label{eqn:PminusLat}                                  
\eea 
In the naive continuum limit, i.e. for infinitesimal $a_-$ \eqnref{eqn:PminusLat} 
becomes the generator of translations along the longitudinal direction. 
However, $P_{-,\mathrm{lat}}$ does not generate translations on the lattice for finite lattice 
spacings. As a consequence, translation invariant states on the lattice are not exact eigenstates of
 $P_{-,\mathrm{lat}}$. There are higher order corrections
in $a_-$ which prevent $P_{-,\mathrm{lat}}$ from being the exact longitudinal lattice translation operator. 

In a numerical simulation with an explicit implementation of the ground state projection operator one has to ensure that the substitution of the lattice Hamiltonian by the effective lattice Hamiltonian is justified. If the relative magnitude of the corrections with respect to the ground state energy is of the order of the time evolution step size $\Delta \tau$ of the Quantum Diffusion Monte Carlo, then the induced defect in the time evolution is effectively of quadratic order in $\Delta \tau$. Hence it can be safely neglected due to the fact that the Quantum Diffusion Monte Carlo algorithm itself is only valid up to quadratic order in $\Delta \tau$. 
In order to quantify the quality of the substitution it is important to measure the typical magnitude of the fluctuations $\langle P_{-,\mathrm{lat}}^2\rangle$ of $P_{-,\mathrm{lat}}$ around its expectation value $\langle P_{-,\mathrm{lat}}\rangle$ where the expectation values are computed with respect to the translation invariant trial/guidance wave functional to which the projection operator is applied.  
Both expectation values are equal to zero for the exact generator of longitudinal translations. 
This is not true for $\mathcal{P}_{-,\mathrm{lat}}$. 
In order to minimize the defect in the time evolution, 
the expectation value of $\mathcal{P}_{-,\mathrm{lat}}$ with respect to the trial wave functional has to be equal to zero and its relative fluctuations have to be of the order of the time evolution step size as discussed.  
Therefore, the trial wave functional has to be selected accordingly.

The effective lattice Hamiltonian can then be chosen as
\bea
\mathcal{H}_{\mathrm{eff,lat}}&=& \frac{1}{N_-N_{\bot}^2}\frac{1}{a_{\bot}^4} \frac{2}{\sqrt{\lambda}}
\sum\limits_{\vec{x}}\left\{
    \frac{1}{2}~\sum\limits_{a}~{\Pi}_-^a(\vec{x})^2~
    +\onehalf~\lambda~\Tr{\unitop-\Real{{U}_{12}(\vec{x})}}
    \right.  \nn\\
& &  \left.
    ~+\sum\limits_{k,a}~\frac{1}{2}~\frac{1}{\tilde{\eta}^2}~ 
      \Bigg[~{\Pi}_k^a(\vec{x})^2+
      \lambda~\Biggl(\Tr{\frac{\sigma_a}{2}~\Imag{U_{-k}(\vec{x})}}\Biggr)^2
      \Bigg] 
    \right\}\; .
\label{eqn:EffectiveLatHamiltonian}    
\eea
By construction, the effective lattice Hamiltonian
\eqnref{eqn:EffectiveLatHamiltonian} is equivalent to a naively
latticized version of the effective continuum Hamiltonian
\eqnref{eq:quadratic}.  For $\tilde{\eta}=1$ this effective lattice
Hamiltonian is very similar to the traditional Hamiltonian used in
equal time lattice theory. They differ in the potential energy terms
for the $U_{-k}$ plaquettes. Instead of the usual
$\mathrm{Tr}[\unitop-\mathrm{Re}(U_{-k})]$ term resembling the field strength squared
in the naive continuum limit, the effective nlc Hamiltonian has the form
 $(\mathrm{Tr}[\sigma^a/2\;\mathrm{Im}(U_{-k})])^2$ which 
corresponds to the plaquette in the adjoint representation. 
These terms which coincide in the continuum
limit have different finite lattice spacing corrections.
\newline
Note that the effective Hamiltonian
\eqnref{eqn:EffectiveLatHamiltonian} has a symmetry which the original
Hamiltonian
\eqnref{eqn:LatticeHamiltonianDensity} did not have, namely it is invariant under a
$Z(2)$ transformation of the following form
\bea
U_k(\vec{x}_\perp,x^-) \rightarrow z\; U_k(\vec{x}_\perp,x^-) \;\;\; \forall \; \vec{x}_\perp\;\mbox{and}\;x^- \; \mbox{fixed} \;,\; z \in Z(2) \;.
\label{eqn:DefSymTrafo}
\eea
Under this transformation, the longitudinal-transversal plaquettes \newline $U_{-k}(\vec{x}_\perp,x^-)$ and  
$U_{-k}(\vec{x}_\perp,x^--1)$ involving transversal links belonging to the longitudinal slice 
$x^-$ transform like
\bea
U_{-k}(\vec{x}_\perp,x^-) &\rightarrow& z \; U_{-k}(\vec{x}_\perp,x^-)\nn\\
U_{-k}(\vec{x}_\perp,x^--1) &\rightarrow& z \; U_{-k}(\vec{x}_\perp,x^--1)\;.
\eea
Of course, this symmetry can be spontaneously broken. In order to
preserve the symmetry properties of the original Hamiltonian we have to
restrict ourselves to the phase  in which the symmetry is spontaneously
broken.  The order parameter of the phase transition is the
expectation value of $\mbox{Tr}\;\mbox{Re}\; U_{-k}$.
In the symmetric phase, the expectation value is equal to zero and in
the broken phase it acquires a non-vanishing expectation value
\begin{equation}
\left\langle \Tr{\Real{ U_{-k}}} \right\rangle \left\{
\begin{array}{l}
= 0     \; \mbox{$Z(2)$ symmetric phase} \\
\neq 0  \; \mbox{$Z(2)$ broken phase}
\end{array}
\right. \;.
\end{equation}
The light cone limit $\tilde{\eta} \rightarrow 0$ 
enhances the importance of transverse
chromo electric and magnetic fields similar to the full nlc Hamiltonian without the unwanted linear terms 
in the momenta. The resulting vacuum solution
should be a plausible extrapolation of
the vacuum solution of  QCD.
     
\section{Analytical solutions of the effective lattice Hamiltonian}
\label{sec:AnalyticalSolutions}

With regard to a subsequent implementation of a guided diffusion 
Monte Carlo it is important to know as much as possible about the true ground
state.   
Analytical solutions of the effective lattice Hamiltonian are possible
in certain regions of the parameter space given by $(\lambda,\eta)$.
In particular, we would like to analyze the
behavior of the ground state wave functional, i.e. the vacuum state, when
the effective parameter $\tilde{\eta} \rightarrow 0$ makes the vacuum approach
the light cone vacuum. Therefore, we have a closer look on the strong ($\lambda<<1$) and 
weak coupling ($\lambda>>1$) solution of the Schr\"odinger equation for the effective lattice 
Hamiltonian in the following. 
  
\subsection{The strong coupling solution of the effective lattice Hamiltonian}
\label{sec:StrongCouplingSolution}
In this section we investigate the strong
coupling limit of the Schr\"odinger equation for which we
are able to find analytic solutions. 
In the strong coupling limit $g>>1$, i.e. $\lambda<<1$ the effective Hamiltonian 
density \eqnref{eqn:EffectiveLatHamiltonian} is dominated 
by chromo electric fields which represent the
kinetic energy terms. 
In comparison with the kinetic energy, the potential energy terms are suppressed by factors 
of $\lambda=4/g^4$.  
Therefore, we may interpret the effective Hamiltonian density as an unperturbed part $\mathcal{T}$
plus a small perturbation $\lambda~\mathcal{V}_{\mathrm{pot}}$ 
\bea
\mathcal{H}_{\mathrm{eff,lat}}&=&\mathcal{T}+\lambda \, \mathcal{V}_{\mathrm{pot}}.
\eea
Here the kinetic energy density $\mathcal{T}$ is given by
\bea
\mathcal{T}&=& \frac{1}{N_-N_{\bot}^2}\frac{1}{a_{\bot}^4} \frac{2}{\sqrt{\lambda}}
               \sum\limits_{\vec{x},a} \left[
   \onehalf \frac{1}{\tilde{\eta}^2} \sum\limits_{k} 
   \Pi_k^a(\vec{x})^2
    +\onehalf \;\Pi_-^a(\vec{x})^2
    \right]\;.
\label{eqn:KineticEnergyOperator}    
\eea
The potential energy density $\lambda~\mathcal{V}_{\mathrm{pot}}$ represents a small
perturbation
\bea
  \mathcal{V}_{\mathrm{pot}} 
&=&\label{eqn:PotEnergy} 
   \frac{1}{N_-N_{\bot}^2}\frac{1}{a_{\bot}^4} \frac{2}{\sqrt{\lambda}}\sum\limits_{\vec{x}} \left\{
   \onehalf \frac{1}{\tilde{\eta}^2} \sum\limits_{k} \left[ 1 - 
      \left( \onehalf \Tr{
         \Real{U_{-k}(\vec{x})}}\right)^2 \right] 
    \right.\nn\\     
    & & \left.
    \phantom{\left(
      \Tr{\frac{\sigma^a}{2}~
         \Imag{U_{-k}(\vec{x})}}\right)^2} 
    +  
    \left[ 1-\onehalf \Tr{\Real{U_{12}(\vec{x})}} 
    \right]
    \right\}\;. \nn\\
& &        
\eea
In order to write the potential energy density $\lambda~\mathcal{V}_{\mathrm{pot}}$ in the 
given form \eqnref{eqn:PotEnergy}, we have used the following identity
\bea
\sum\limits_{a} \left(\Tr{\frac{\sigma^a}{2}~\Imag{U_{-k}(\vec{x})}}\right)^2 
&=& 1 - \left( \onehalf \Tr{\Real{U_{-k}(\vec{x})}}\right)^2 \; .
\eea
We perform perturbation theory in $\lambda$. Then the ground state $\ket{\Psi_0}$
as well as the ground state energy density $\epsilon_0$
are written as a power series in $\lambda$ where ``$(n)$" denotes the $n$-th order 
correction
\bea
\ket{\Psi_0} &=& \sum\limits_{n=0}^{\infty}\lambda^n \ket{\Psi_0^{(n)}} \nn\\
\epsilon_0&=& \sum\limits_{n=0}^{\infty} \lambda^n \epsilon_0^{(n)}\;.
\eea
The unperturbed Hamiltonian $\mathcal{T}$ is a sum of quantum rigid rotators, 
one for each lattice site and for each spatial direction \cite{Kogut:1974ag}. The spectrum of each $\sum_a\Pi^{a\,2}$ is given by $E_{l}=l(l+1)$ with $l\in (0,1/2,1,\ldots)\;$ in $SU(2)$. 
Each eigenvalue $E_l$ is $(2l+1)^2$-fold degenerate.
Therefore, the unperturbed ground state $|\Psi_0^{(0)}\rangle$ of $\mathcal{T}$ is the state which has $l=0$ for each rotator. It is annihilated by all the momentum operators 
\bea
\Pi_j^a(\vec{x}) \ket{\Psi_0^{(0)}}= 0\;\;\; \forall\;\vec{x},a\; \wedge \; \forall \;j\in\{1,2,-\}\;.
\eea
This state does not depend on the $U_j(\vec{x})$ in the link-coordinate
representation, i.e. is a constant and is
non-degenerate. The corresponding ground state energy is given by 
\bea
\epsilon_0^{(0)}=0.  
\eea 
The space of states may be constructed from the ground state $|\Psi_0^{(0)}\rangle$ by
applying the link operator in a given representation $(l)$ which is then again an eigenstate
of $\sum\limits_a \Pi_j^a(\vec{x})^2$ with eigenvalue $E_l$
\bea
\sum\limits_a \Pi_j^a(\vec{x})^2 ~ U_j^{(l)}(\vec{x}) ~ \ket{\Psi_0^{(0)}} &=&
l(l+1)~ U_j^{(l)}(\vec{x}) ~ \ket{\Psi_0^{(0)}}~.
\eea    
Note that the representation index $(l)$ of the link explicitly refers to its
$SU(2)$-representation whereas links without a representation index are defined to be in the 
fundamental representation 
\bea
U_j(\vec{x})\equiv U_j^{(1/2)}(\vec{x}) \;.
\eea
Due to the non-degenerate ground state we may apply standard 
Raleigh-Schr\"odinger perturbation theory.
In general,
the first order correction to the ground state reads 
\bea
\ket{\Psi_0^{(1)}}&=&
\frac{1}{\epsilon_0^{(0)}-\mathcal{T}}\mathcal{V}_{\mathrm{pot}}\ket{\Psi_0^{(0)}}
\nn\\ &=&
-\frac{1}{\mathcal{T}}\mathcal{V}_{\mathrm{pot}}\ket{\Psi_0^{(0)}}\;.
\label{eqn:FirstOrderCorrection}
\eea
The correspondent first order correction to the ground state energy density is given by
\bea
\epsilon_0^{(1)} &=& \bra{\Psi_0^{(0)}} \mathcal{V}_{\mathrm{pot}} \ket{\Psi_0^{(0)}} \;.
\eea
It is a Haar integral over the whole configuration space which is given by
\bea
\epsilon_0^{(1)} &=& \int \mathcal{V}_{\mathrm{pot}}\left(U\right)
                     \prod\limits_{\vec{x},j} dU_{j}(\vec{x}) \nn\\
                 &=& \frac{1}{\tilde{\eta}^2}
                     \frac{1}{a_{\bot}^4} \frac{2}{\sqrt{\lambda}}
                     \left(\frac{3}{4}+\tilde{\eta}^2\right)\lambda\;.
\eea
This yields a total ground state energy density in the strong coupling limit 
\bea
\epsilon_0 &=& \frac{1}{\tilde{\eta}^2}
               \frac{1}{a_{\bot}^4}\left[ 
               \left(\frac{3}{2}+2~\tilde{\eta}^2\right)\sqrt{\lambda}+\mathcal{O}(\lambda^{3/2})
               \right]\; .
\eea
In order to compute \eqnref{eqn:FirstOrderCorrection} we use the fact that the
trace of the plaquette $U_{12}(\vec{x})$ and the squared trace of the plaquette $U_{-k}(\vec{x})$ minus one are eigenstates of the kinetic energy operator with eigenvalues $t_{-}$ and
$t_{\bot}$, respectively
(cf. \eqnref{eqn:DoubleCommTrRe} and \eqnref{eqn:DoubleCommTrRe2} in
appendix \ref{app:CommRel})
\bea
\mathcal{T} ~ \Tr{\Real{U_{12}(\vec{x})}}  \ket{\Psi_0^{(0)}} &=& t_{-} ~ \Tr{\Real{U_{12}(\vec{x})}} ~ \ket{\Psi_0^{(0)}} \nn\\
\mathcal{T} \; \left[\left(\Tr{\Real{U_{-k}(\vec{x})}}\right)^2-1\right] \ket{\Psi_0^{(0)}}
&=& \nn\\ 
& & \hspace{-3.0cm} t_{\bot} \left[\left(\Tr{\Real{U_{-k}(\vec{x})}}\right)^2-1\right] \ket{\Psi_0^{(0)}}\;.
\eea
The eigenvalues $t_-$ and $t_\bot$ are given by
\bea
t_{-} &=& \left[ \frac{1}{N_-N_{\bot}^2}\frac{1}{a_{\bot}^4} \frac{2}{\sqrt{\lambda}}  \frac{2}{\tilde{\eta}^2}\right]\cdot \frac{3}{4} \nn\\
t_{\bot} &=& \left[\frac{1}{N_-N_{\bot}^2}\frac{1}{a_{\bot}^4}\frac{2}{\sqrt{\lambda}}  \left(1+ \frac{1}{\tilde{\eta}^2}\right)\right]\cdot 2 \;.
\eea
The factor $3/4$ in $t_{-}$ is related to the fundamental representation ($l=1/2$) of the plaquette and the factor of $2$ in $t_{\bot}$ arises from the squared trace of the plaquette minus one in the fundamental representation which is equivalent to the trace of the plaquette in the adjoint representation ($l=1$).
Hence, the first order
correction to the ground state wave functional is given by 
\bea
\ket{\Psi_0^{(1)}}&=& \sum\limits_{\vec{x}} \left\{
\frac{1}{3}~\tilde{\eta}^2~\Tr{\Real{U_{12}(\vec{x})}} \right.\nn\\ &
& \left.\hspace{0.1\linewidth}+
\frac{1}{16}~\frac{1}{1+\tilde{\eta}^2}~\sum\limits_k
\left(\Tr{\Real{U_{-k}(\vec{x})}}\right)^2 \right\} \ket{\Psi_0^{(0)}}\;.\nn\\
& &
\eea 
The state $|\Psi_0^{(1)}\rangle$ does not contain any products of plaquettes involving
field strengths at different spatial positions. Therefore, to this order in perturbation theory, the ground state
wave functional factorizes in a product of single plaquette
wave functionals similar to the vacuum wave functional obtained for an
equal time lattice Hamiltonian \cite{Chin:1985ua}
\bea 
\label{eqn:StrongCouplingSolution}
\ket{\Psi_0} &=& \left\{\unitop
+ \lambda \sum\limits_{\vec{x}} \left[ \frac{1}{3}\tilde{\eta}^2
\Tr{\Real{U_{12}(\vec{x})}} \right.\right.\nn\\ & & \left.\left.+
\frac{1}{16}\,\frac{1}{1+\tilde{\eta}^2}\sum\limits_k
\left(\Tr{\Real{U_{-k}(\vec{x})}}\right)^2
\right]+\mathcal{O}(\lambda^2) \right\}\ket{\Psi_0^{(0)}} \nn\\ &=&
\prod\limits_{\vec{x}} \exp\Bigg\{ \frac{1}{3}~\lambda~\tilde{\eta}^2
\Tr{\Real{U_{12}(\vec{x})}} \Bigg. \label{eqn:FOSCSol}\\ & &
\Bigg. \hspace{0.05\linewidth}
+\frac{1}{16}\,\frac{\lambda}{1+\tilde{\eta}^2}\sum\limits_k
\left(\Tr{\Real{U_{-k}(\vec{x})}}\right)^2 
\Bigg\}\ket{\Psi_0^{(0)}}+\mathcal{O}(\lambda^2) 
 \;.\nn
\eea
In the wave functional, the purely transversal plaquettes $U_{12}$ involving the 
longitudinal chromo magnetic fields are suppressed by $\tilde{\eta}^2$ in the light
cone limit $\tilde{\eta} \rightarrow 0$. 
To this order in perturbation theory, the strong coupling ground state wave functional
\eqnref{eqn:StrongCouplingSolution} respects the $Z(2)$ symmetry of the effective 
Hamiltonian which the full Hamiltonian, however, does not share.

\subsection{Weak coupling solution of the effective lattice Hamiltonian}
\label{sec:WeakCouplingSolution}

In the weak coupling regime, i.e. $g\rightarrow
0$ or $\lambda \rightarrow \infty$ the effective lattice
Hamiltonian \eqnref{eqn:EffectiveLatHamiltonian} in $SU(2)$ depends on a
triplet of free $U(1)$ gauge fields and their 
corresponding momenta. To reduce the Hamiltonian into this form it is convenient to
substitute the gauge field $g \,A_i^a(\vec{x})$ in \eqnref{eqn:PrimDefLink} by a rescaled 
gauge field $\widetilde{A}_i^a(\vec{x})$ (cf. \eqnref{eqn:DefRescaledFields}). 
Note that all vector indices throughout this section
refer to a flat space metric equal to the unit matrix. Furthermore,  
$\epsilon_{ijk}$ is the totally antisymmetric Levi-Cevita symbol
with $\epsilon_{12-}=1$. In the $g\rightarrow0$ limit, the field strength tensor reduces to the chromo magnetic field 
$B_i^a(\vec{x})$, which is the $i$-th spatial component of
the lattice curl of $\vec{A}^a\left(\vec{x}\right)$ and which is 
rescaled to $\widetilde{B}_i^a(\vec{x})$
\bea
g\,A_i^a(\vec{x}) &=& \frac{\widetilde{A}_i^a(\vec{x})}{\sqrt{\lambda}} 
\;\;\;\;\;\;\;\;\;\; i=1,2,-\nonumber\\
g\,B_i^a(\vec{x})  &=& g\, \epsilon_{ilm} \left[A_m^a(\vec{x})-A_m^a(\vec{x}-\vec{e}_l)\right]\nn\\
g\,B_i^a(\vec{x})&=&\frac{\widetilde{B}_i^a(\vec{x})}{\sqrt{\lambda}}.
\label{eqn:DefRescaledFields}
\eea
Similarly to the equal
time theory \cite{Chin:1983pc,Chin:1985ua} one can expand the effective lattice
Hamiltonian in a power series in $\lambda^{-1}$. The expansion of the potential energy is
straightforward. The kinetic energy of the effective lattice Hamiltonian is a sum of the 
Casimir operators acting on $SU(2)$. Each of them represents a Laplace-Beltrami 
operator on the curved manifold of $SU(2)$. The expansion in a power series of this operator 
yields in leading order a flat space Laplacian in three dimensions given by
\bea
\sum\limits_a \widetilde{\Pi}_j^a(\vec{x})^2 &=& -\sum\limits_a \frac{\delta^2}{\delta \widetilde{A}_j^a(\vec{x})^2}\;.
\eea  
Hence, the $\tilde{\Pi}^a$,$\tilde{A}^a$ obey effectively the following commutation relations
\bea
\left[\widetilde{\Pi}_i^a(\vec{x}),\widetilde{A}_j^b(\vec{y})\right]&=&
-\mathrm{i}\, \delta_{ab}\delta_{ij}\delta_{\vec{x},\vec{y}} \nn\\
\left[\widetilde{\Pi}_i^a(\vec{x}),\widetilde{\Pi}_j^b(\vec{y})\right]&=&
0\nn\\
\left[\widetilde{A}_i^a(\vec{x}),\widetilde{A}_j^b(\vec{y})\right]&=&
0.  
\eea
The described expansion of the effective lattice Hamiltonian in the weak coupling limit yields
in leading order
\bea
\mathcal{H}_{\mathrm{eff,lat}}&=& \frac{1}
{N_-N_{\bot}^2}\frac{1}{a_{\bot}^4} \frac{1}{\sqrt{\lambda}}
\sum\limits_{\vec{x},a}\left\{
    \lambda~\widetilde{\Pi}_-^a(\vec{x})^2~
    +\frac{1}{4}~\widetilde{B}_-^a(x)^2 
    \right.  \nn\\
& &  \left.
    ~+\sum\limits_{k}~\frac{1}{\tilde{\eta}^2}~ 
      \Bigg[\lambda~\widetilde{\Pi}_k^a(\vec{x})^2+\frac{1}{4}~\widetilde{B}_k^a(x)^2
      \Bigg] 
    \right\}+\mathcal{O}\left(\frac{1}{\tilde{\eta}^2\lambda^{5/4}}\right)\; .\nn\\
& &    \label{eqn:WeakCouplingHamiltonian}
\eea
This Hamiltonian is equivalent to the abelian limit and
the order $\lambda^{-5/4}$ corrections represent the triple gluon vertex $g\,AAA$. 

\begin{figure}[t]
\centering	
	\begin{minipage}[t]{0.8\textwidth}
    \includegraphics[width=1.00\textwidth]{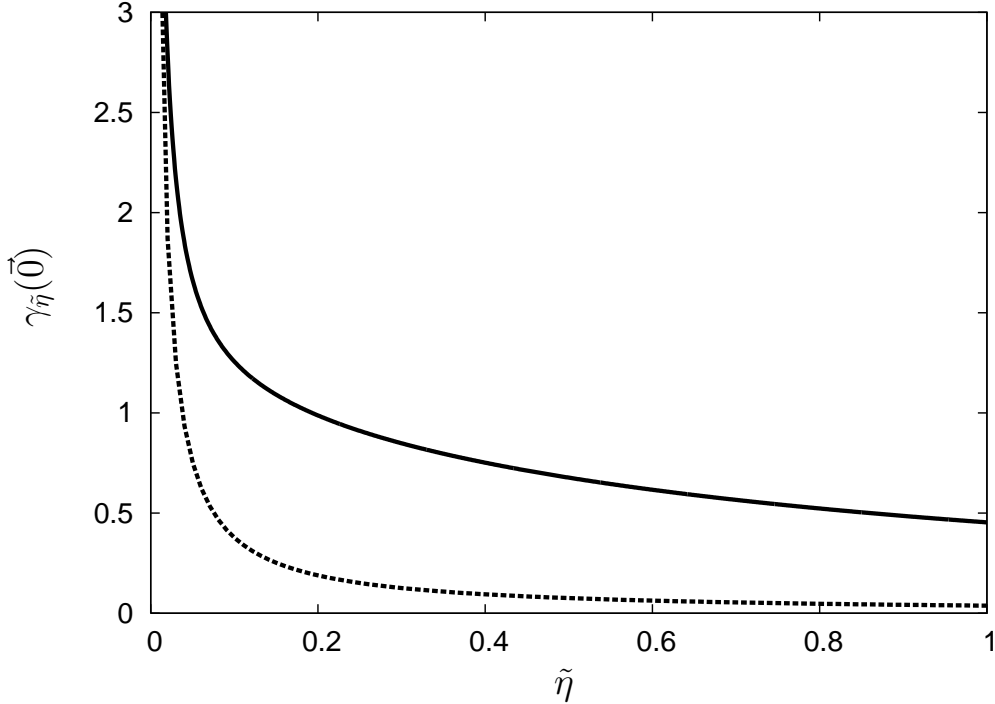}
	\end{minipage}
	\centering
	\caption{\label{fig:Gammaeta(0)_16x16x16}              
	         Diagonal element $\gamma_{\tilde{\eta}}(\vec{0})$ of the spatial part of the 
	         covariance matrix for a $16^3$-lattice as a function of
	         $\tilde{\eta}$ (solid line). Its asymptotic behavior in the light cone limit
$\tilde{\eta} \rightarrow 0$, $\gamma_{\tilde{\eta}}(\vec{0})\sim 0.038/\tilde{\eta}$
	         is shown by the dashed line. 
	        }
\end{figure}

Instead of solving the ground state
in terms of the gauge variant fields $\widetilde{A}_k$ as done 
in ref. \cite{Hamer:1993db}, we express the kinetic energy
operator acting on the gauge fields in terms of effective operators
which act on chromo magnetic fields $\widetilde{B}_k$. These are gauge invariant in the abelian
limit. By doing so, we 
obtain a ground state wave functional which depends only on gauge
invariant objects. This avoids an otherwise necessary projection onto
a gauge invariant subspace of the Hilbert space.  After
transforming the Hamiltonian into Fourier space, several unitary
transformations convert the Hamiltonian into a
Hamiltonian of decoupled harmonic oscillators. The
necessary unitary transformations are
similar to transformations performed for a compact equal time
$U(1)$ Hamiltonian in ref.~\cite{Hamer:1993db}. However, additional factors
due to the nlc metric appear which can be traced in the computation. Once
the harmonic oscillator Hamiltonian is obtained, the ground state wave functional
and the ground state energy $\epsilon_0 $ are easily found
\bea
\epsilon_0 &=&\frac{1}{a_{\bot}^4}  \frac{6}{\tilde{\eta}^2} 
\frac{1}{N_-N_{\bot}^2} \sum\limits_{\vec{k}} 
\left[\tilde{\eta}^2 \sin\left(\frac{k_1}{2}\right)^2+
\tilde{\eta}^2 \sin\left(\frac{k_2}{2}\right)^2+ \sin\left(\frac{k_-}{2}\right)^2 \right]^{1/2} \; . \nn\\
& & 
\label{eqn:EDensWC}
\eea
Here $k_i$ denote the lattice momentum values
\bea
k_i \equiv \frac{2 \pi}{N_i} n_i \;\;\;\; n_i=0,\ldots,N_i-1 \;.
\eea
\begin{figure}[t]
	\centering
	\begin{minipage}[t]{0.8\textwidth}
   	\includegraphics[width=1.00\textwidth]{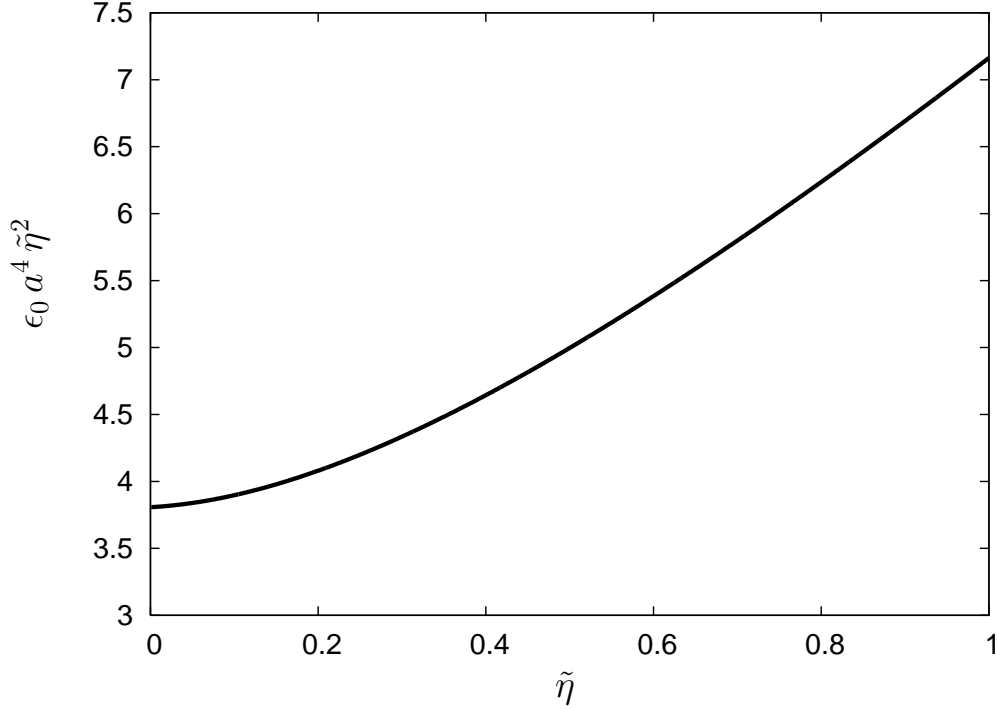}
	\end{minipage}
	\caption{\label{fig:Gammaepsilon_16x16x16}
	         Rescaled dimensionless energy density $\epsilon_0 a^4 \tilde{\eta}^2$ of the effective 
	         nlc Hamiltonian in leading order of the weak coupling limit for a $16^3$-lattice 
	         as a function of $\tilde{\eta}$
	        }
\end{figure}
In \figref{fig:Gammaepsilon_16x16x16}
we show the dimensionless energy density \eqnref{eqn:EDensWC}
for a $16^3$-lattice as a
function of $\widetilde{\eta}$. A leading $1/\tilde{\eta}^2$
-dependence of the effective energy density is
obvious from \eqnref{eqn:EDensWC} and arises from the $1/\tilde{\eta}$ dependence
of the light cone energy and the $\tilde{\eta}$ dependence of the volume $V=N_{\perp}^2N_- a_\perp^2a_-$.
This dependence is scaled out in the figure.  In the
abelian limit, the energy density is given by the 
dispersion relation summed over all modes, times the color degeneracy
factor. If we identify $p_i=\sin(\pi \, n_i/N_i)$ with the
latticized version of the $i$-th component of the gluon momentum
$p_i$, then the nlc dispersion relation $\omega_{nlc}$ of a free gluon gas
is given by (cf. \eqnref{eqn:MomentaRel})
\bea
\omega_{nlc} &=& \frac{1}{\tilde{\eta}}\left.
\Big(\sqrt{p_1^2+p_2^2+p_3^2}-p_3\Big)\right|_{p_3=p_-/\tilde{\eta}}\;.
\eea
Here $p_3$ refers to the longitudinal mode in the laboratory frame
and $p_-$ refers to the longitudinal mode in the nlc frame.  Hence, by
summing $\omega_{nlc}$ over all modes and taking into account that the total
longitudinal momentum adds up to $P_-=0$ we obtain
\eqnref{eqn:EDensWC}.
The ground state wave functional  is a multivariate Gaussian wave functional in the chromo magnetic fields
where $\Gamma_{\tilde{\eta}}^{ij}(\vec{x}-\vec{x}')$ denote the matrix
elements of the covariance matrix
\bea
\Psi_0&=&\exp\left\{-\sqrt{\lambda}
                     \sum\limits_{\vec{x},\vec{x}'} \sum_{a,i,j}
                     \frac{g}{2} B_i^a(\vec{x}) \Gamma_{\tilde{\eta}}^{ij}(\vec{x}-\vec{x}')
                     \frac{g}{2} B_j^a(\vec{x}')
\right\}
\label{eqn:WeakCouplingSolution} \nonumber\\
\mathbf{\Gamma}_{\tilde{\eta}}(\vec{x}-\vec{x}')&\equiv&
\left(
\begin{array}{ccc}
\gamma_{\tilde{\eta}}(\vec{x}-\vec{x}')&0&0\\
0&\gamma_{\tilde{\eta}}(\vec{x}-\vec{x}')&0\\
0&0&\tilde{\eta}^2\gamma_{\tilde{\eta}}(\vec{x}-\vec{x}')
\end{array}
\right)\;.
\label{eqn:CovMatrix}
\eea
Here $\gamma_{\tilde{\eta}}$ denotes the spatial part of the covariance matrix.
It depends only
on the relative distance $\vec{x}-\vec{x}'$ of the
chromo magnetic fields in the wave functional
\bea
\lefteqn{
\gamma_{\tilde{\eta}}(\vec{x}-\vec{x}')\equiv  
\onehalf \frac{1}{N_-N_{\bot}^2} } & & \label{eqn:Gamma}\\
& & \times
\sum\limits_{\vec{k}\ne\vec{0}}
\left[\tilde{\eta}^2 \sin\left(k_1/2\right)^2+\tilde{\eta}^2 \sin\left(k_2/2\right)^2+ \sin\left(k_-/2\right)^2 \right]^{-1/2}e^{\mathrm{i}\vec{k}\cdot\left(\vec{x}-\vec{x}'\right)}\;.
\nn
\eea
The function $\gamma_{\tilde{\eta}}(\vec{x}-\vec{x}^\prime)$ is real due to the invariance  under space reflections.
In \figref{fig:Gammaeta(0)_16x16x16}
we show $\gamma_{\tilde{\eta}}(\vec{0})$ for a $16^3$-lattice as
a function of $\tilde{\eta}$.
The asymptotic behavior of $\gamma_{\tilde{\eta}}(\vec{0})$ 
 in the light cone limit 
$\tilde{\eta}\rightarrow 0$ can be computed by summing all modes with $\vec{k}\neq \vec{0}$ and
$k_-=0$ in \eqnref{eqn:Gamma}.
For a $16^3$-lattice, it is given by
\bea
\gamma_{\tilde{\eta}}(\vec{0}) 
\sim  \frac{0.038}{\tilde{\eta}} \;\;,\;\;\tilde{\eta} \rightarrow 0 \;.
\eea
For $\tilde{\eta}=1$ the $3\times3$ covariance matrix $\mathbf{\Gamma}_{\tilde{\eta}}$ 
\eqnref{eqn:CovMatrix} equals the covariance matrix which was found by Chin et al.
\cite{Chin:1983pc,Chin:1985ua} for an equal time theory 
since our Hamiltonian coincides with the
correspondent equal time Hamiltonian in the weak coupling limit.
For small values of
$\tilde{\eta}$ the chromo magnetic field in the longitudinal direction
$B_-^a \propto F_{12}^a$ is suppressed in the wave functional by a factor 
of $\tilde{\eta}$ in comparison with the other field strengths.  
On the other hand, the
chromo magnetic fields in transversal directions $B_1^a \propto F_{-2}$ and $B_2^a \propto F_{-1}$ 
are not suppressed.

\begin{figure}[!h]
	\centering
	\begin{minipage}[t]{0.6\textwidth}
		\centering
		\includegraphics[width=1.00\textwidth]{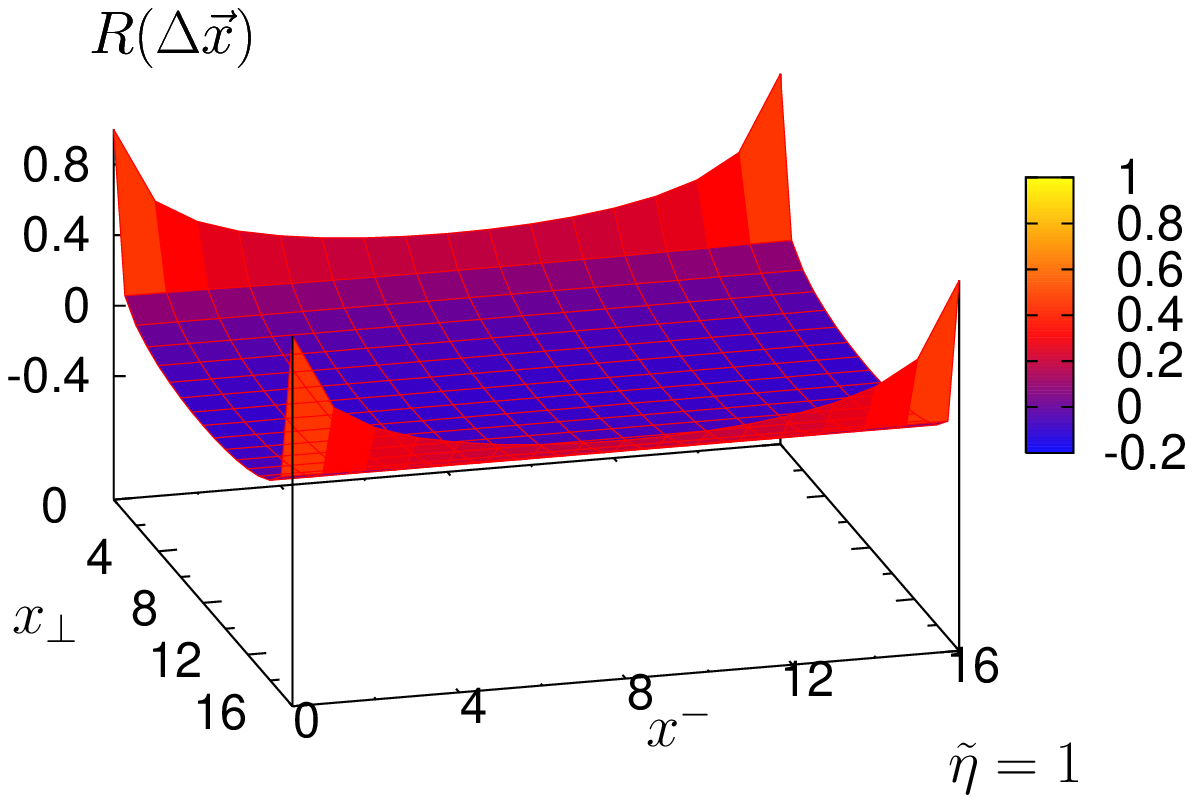}
	\end{minipage}
	\begin{minipage}[t]{0.6\textwidth}
		\centering
    \includegraphics[width=1.00\textwidth]{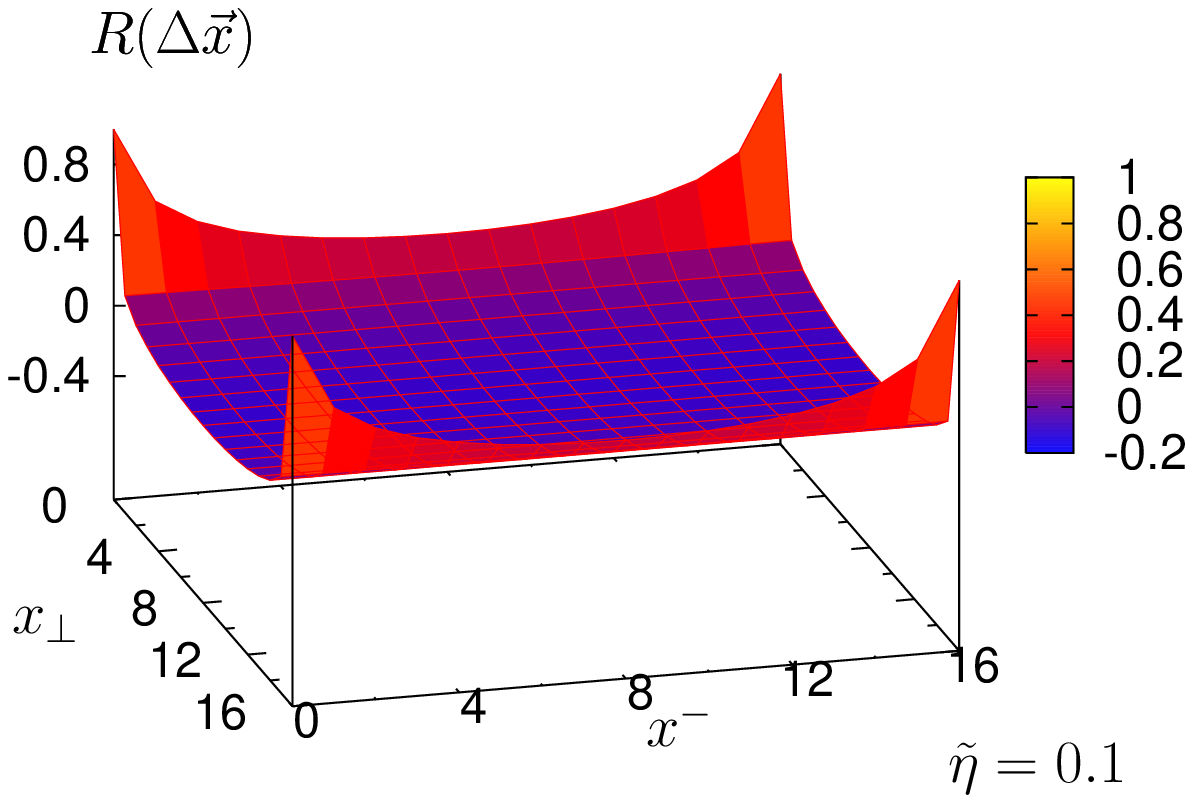}
	\end{minipage}
	\centering
	\begin{minipage}[b]{0.6\textwidth}
		\centering
    \includegraphics[width=1.00\textwidth]{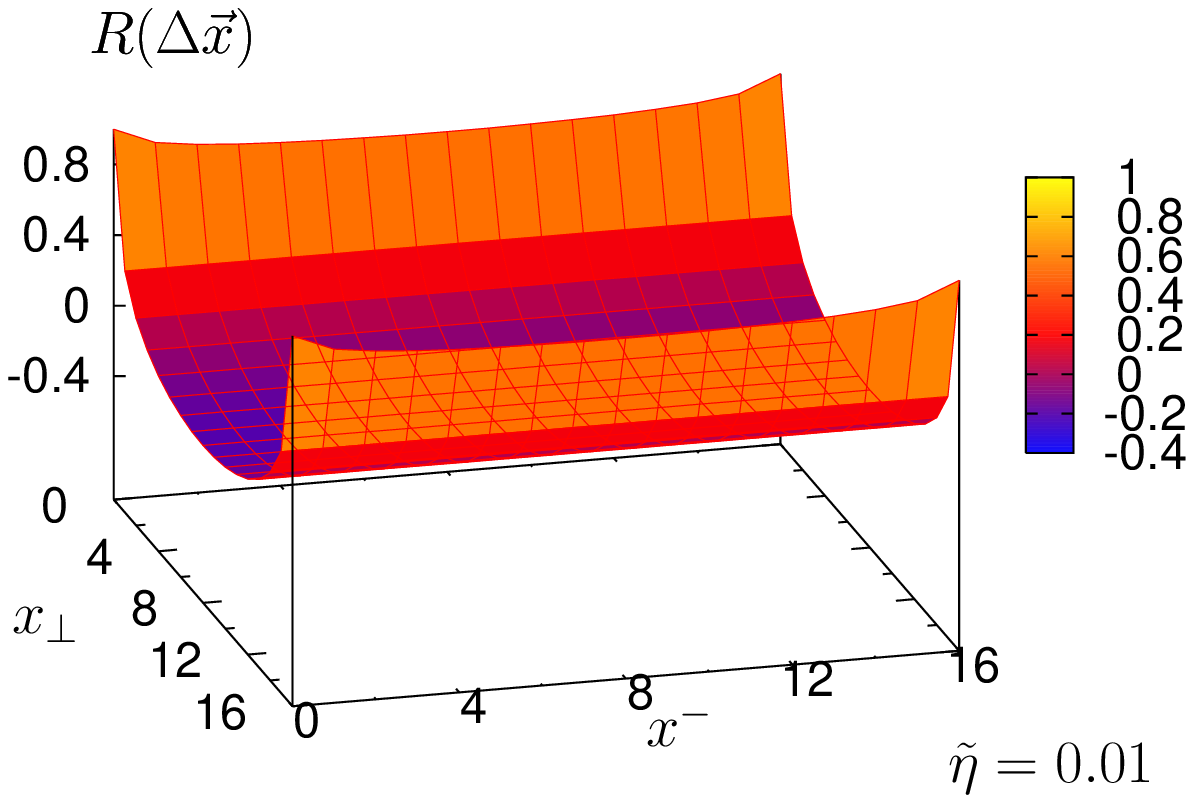}
	\end{minipage}
	\centering
	\caption{Ratio of covariance matrix elements $R(\Delta\vec{x})$ as a function of the separation 
	         $\Delta \vec{x}$ for a 2 dimensional $16\mathrm{x}16$ lattice at three different values 
	         of $\eta^2$.\label{fig:ConnCoeff}}
\end{figure}

We compare correlation matrix elements $\gamma_{\tilde{\eta}}(\vec{x}-\vec{x}^\prime)$ 
for $\Delta \vec{x} \neq 0$ with the matrix element at $\Delta \vec{x} = 0$ by forming 
the ratio $R(\Delta\vec{x})$
\bea
R(\Delta\vec{x})&\equiv&\frac{\gamma_{\tilde{\eta}}(\Delta\vec{x})}{\gamma_{\tilde{\eta}}(\vec{0})}.
\eea
In \figref{fig:ConnCoeff}, $R(\Delta\vec{x})$ is 
shown for a $16 \mathrm{x} 16$-lattice and 
for three different values of $\tilde{\eta}$, 
namely $\tilde{\eta}=1$,$\tilde{\eta}=10^{-1}$ and $\tilde{\eta}=10^{-2}$. 
For reasons of presentability, we have restricted ourselves to a
$2$-dimensional section through the $3$-dimensional lattice spanned by
$x_\bot=x^1$ and $x^-$ at $x^2=0$. This representation allows to see
the anisotropy developing for very small $\tilde{\eta}$.  
Here and in the following, the notion ``off-diagonal in position space" refers to $\Delta
\vec{x}\neq 0$ whereas ``diagonal in position space" refers to
$\Delta \vec{x}=0$. 
For $\tilde{\eta}=1$, the covariance matrix has only
weakly-off diagonal contributions in position space. 
Therefore, it is reasonable to consider the weak
coupling wave functional in the diagonal approximation
\cite{Chin:1985ua} as a product of single plaquette functionals.
For decreasing $\tilde{\eta}$ one observes that
the correlations among plaquettes separated along the longitudinal
direction become more and more important. In the light cone limit,
every plaquette is equally correlated with any other plaquette which
is longitudinally separated from the first one.

However, for not too small values of $\tilde{\eta}$, at least an
effective description by a product of single plaquette wave functionals
is possible. In \secref{sec:OutlookAndConclusions} we discuss a possibility 
to include long range correlations in the wave functional by a combined
optimization and Quantum Diffusion Monte Carlo method. 

\FloatBarrier

\section{Variational optimization of the ground state wave functional}
\label{sec:VarOpt}
In the last two sections we have analyzed the strong and weak coupling behavior of 
the Hamiltonian and its ground state. We have seen that  
in the strong coupling limit the ground state wave functional may
be approximated by a product of single site wave functionals.
Also in the weak coupling limit for not too small $\tilde{\eta}$ the
bilocality of the chromo magnetic field strength is less important. In the 
following we construct an effective wave functional which smoothly
interpolates between the strong and weak coupling solution.  
In addition we would like to 
choose the ground state wave functional in such a way that it is 
not invariant under the unwanted additional $Z(2)$ symmetry of the effective
Hamiltonian
in which the linear momentum terms are compensated by the translation
operator. 
Therefore, me make a variational ansatz of the ground state wave functional for the
whole coupling range which contains a product of single site
plaquettes with two variational parameters $\rho$ and $\delta$. We denote the normalization
constant by N
\bea
\Psi_0(\rho,\delta)&=& N \prod\limits_{\vec{x}}\exp\left\{\sum\limits_{k=1}^2 
                 \rho\,\Tr{\Real{U_{-k}(\vec{x})}}
                +\delta\,\Tr{\Real{U_{12}(\vec{x})}}
               \right\}\;. \nn\\
           & & \phantom{1}    
\label{eqn:VariationlGSWF}
\eea
With this normalized wave functional we variationally optimize the 
energy expectation value $\epsilon_0(\rho,\delta)$ of the effective 
Hamiltonian which is given 
in terms of plaquette expectation values.
\bea
\epsilon_0(\rho,\delta)&=&\bra{\Psi_0}\mathcal{H}_{eff}\ket{\Psi_0}\nonumber\\
               &=&\phantom{+}\frac{1}{N_-N_{\bot}^2}\frac{1}{a_{\bot}^4} \frac{2}{\sqrt{\lambda}}
                      \sum\limits_{\vec{x}\phantom{,k}}
                      \left[\left( \frac{3}{4} \frac{\delta}{\tilde{\eta}^2} 
                            -\frac{\lambda}{2} \right)
                      \left\langle \Tr{\Real{U_{12}(\vec{x})}} \right\rangle +\lambda\right] 
                      \nn\\
               & &+\frac{1}{N_-N_{\bot}^2}\frac{1}{a_{\bot}^4} \frac{2}{\sqrt{\lambda}}
                    \sum\limits_{\vec{x},k}\left[\frac{3}{8}\rho\left( 1 + \frac{1}{\tilde{\eta}^2}\right)
                      \left\langle \Tr{\Real{U_{-k}(\vec{x})}} \right\rangle 
                      \right]
                      \nn\\                    
               & &+\frac{1}{N_-N_{\bot}^2}\frac{1}{a_{\bot}^4} \frac{2}{\sqrt{\lambda}}
                    \sum\limits_{\vec{x},k}\left[\frac{\lambda}{2}\frac{1}{\tilde{\eta}^2}
                      \left(1-\frac{1}{4}\left\langle \left(\Tr{\Real{U_{-k}(\vec{x})}}\right)^2 \right\rangle \right)
                      \right] \;. \nn\\
               & &        
\label{eqn:ExpValEnergyDens}
\eea
The explicit dependence of the energy expectation value on $\rho$ and $\delta$
comes from the kinetic energy terms in $\mathcal{H}_{eff}$. There is an implicit
dependence in the plaquette expectation values which are computed as averages over 
link configurations generated by the probability density
\bea
dP(U)&=& \left|\Psi_0(\rho,\delta) \right|^2 \prod\limits_{\vec{x},j} \mathcal{D}U_j(\vec{x})~.
\label{eqn:ProbDistrib}
\eea 
With the special form of our trial ground state
wave functional \eqnref{eqn:VariationlGSWF}, the energy expectation
value of the effective Hamiltonian
\eqnref{eqn:EffectiveLatHamiltonian} coincides with the energy
expectation value of the full Hamiltonian
\eqnref{eqn:LatticeHamiltonianDensity}. Even if we do not use the 
invariance of the trial wave functional under translations, the
expectation value of the longitudinal momentum operator with respect
to the trial wave functional \eqnref{eqn:VariationlGSWF} vanishes
identically. This is due to the fact that the 
expectation value of the chromo electric field operator $\Pi_j^a(\vec{y})$
times an arbitrary functional $G(\{U\})$ of the links with respect to a
purely real valued exponential wave functional obeys  
\bea
\bra{\Psi_0} \Pi_j^a(\vec{y})~G(\{U\})\ket{\Psi_0} &=&-\bra{\Psi_0}
G(\{U\})~\Pi_j^a(\vec{y})\ket{\Psi_0} \; .
\label{eqn:partialint}  
\eea
The above relation \eqnref{eqn:partialint} may be interpreted as a ``partial" 
integration rule and is proven in appendix \ref{app:CommRel}. Hence, the 
ground state wave functional \eqnref{eqn:VariationlGSWF} minimizing the
energy density \eqnref{eqn:ExpValEnergyDens}
optimizes simultaneously the effective and the full Hamiltonian. 
In order to optimize the ground state wave functional we
sample the probability distribution \eqnref{eqn:ProbDistrib} with a
local heat bath algorithm \cite{Creutz:1980zw} on a $16^3$-lattice and
measure the expectation values of the plaquettes and the squared
plaquettes with the bootstrap method \cite{Efron:1979} using an initial sample size of 
$500$ and a bootstrap sample size of $1000$. We compute these expectation values
as a function of the parameters $\rho$
and $\delta$ on a $50 \times 50 $ grid where each of the parameters varies in the
interval $[0,10]$ with a sterilize of $0.2$. This yields a set of
$2500$ distinct expectation values which we interpolate
with polynomials of fifth order.
For a first coarse estimate of the optimized parameters, we find 
the minimum of \eqnref{eqn:ExpValEnergyDens} with a standard Mathematica
minimization routine. 

For the fine determination of the optimal parameters we
then generate 50 different pairs with energy 
expectation values less than three percent higher
than the energy at the coarse estimate of $\rho_0,\delta_0$. 
Finally, we fit these energy values with 
a quadratic form \eqnref{eqn:quadraticform}
centered at the optimal values $(\rho_0,\delta_0)$ where
the linear term in the taylor series vanishes due to the minimum
condition
\begin{equation}
\epsilon_0(\rho,\delta)\approx \epsilon_0(\rho_0,\delta_0)+\onehalf
\left(
\begin{array}{c}
\rho-\rho_0 \\
\delta-\delta_0 
\end{array}
\right)^\mathrm{T}
\cdot 
\left(
\begin{array}{cc}
\frac{\partial^2 \epsilon_0}
{\partial_\rho \partial_\rho} & \frac{\partial^2 \epsilon_0}{\partial_\rho \partial_\delta} \\
\frac{\partial^2 \epsilon_0}
{\partial_\delta \partial_\rho} & \frac{\partial^2 \epsilon_0}{\partial_\delta \partial_\delta}
\end{array}
\right) 
\cdot
\left(
\begin{array}{c}
\rho-\rho_0 \\
\delta-\delta_0
\end{array}
\right)\;.
\label{eqn:quadraticform}
\end{equation}
The described method is tested by comparing our results with the
variational results of Chin et al.  \cite{Chin:1985ua} who optimized
a one parameter ($2\,\rho=2\,\delta\equiv\alpha_\mathrm{Chin}$) wave functional of
the form given in \eqnref{eqn:VariationlGSWF} with respect to the
standard equal time Hamiltonian containing only plaquette terms
without anisotropy. Note that Chin's results have been
obtained on a $4^3$-lattice, but the authors show that the
dependence of the energy density and the optimal wave functional
parameter on the lattice size is small.   We find $0.5\%$ 
agreement between the results of
our method and the values obtained by Chin et
al. \cite{Chin:1985ua}. 

\begin{figure}
	\centering
	    \includegraphics[width=0.9\textwidth]{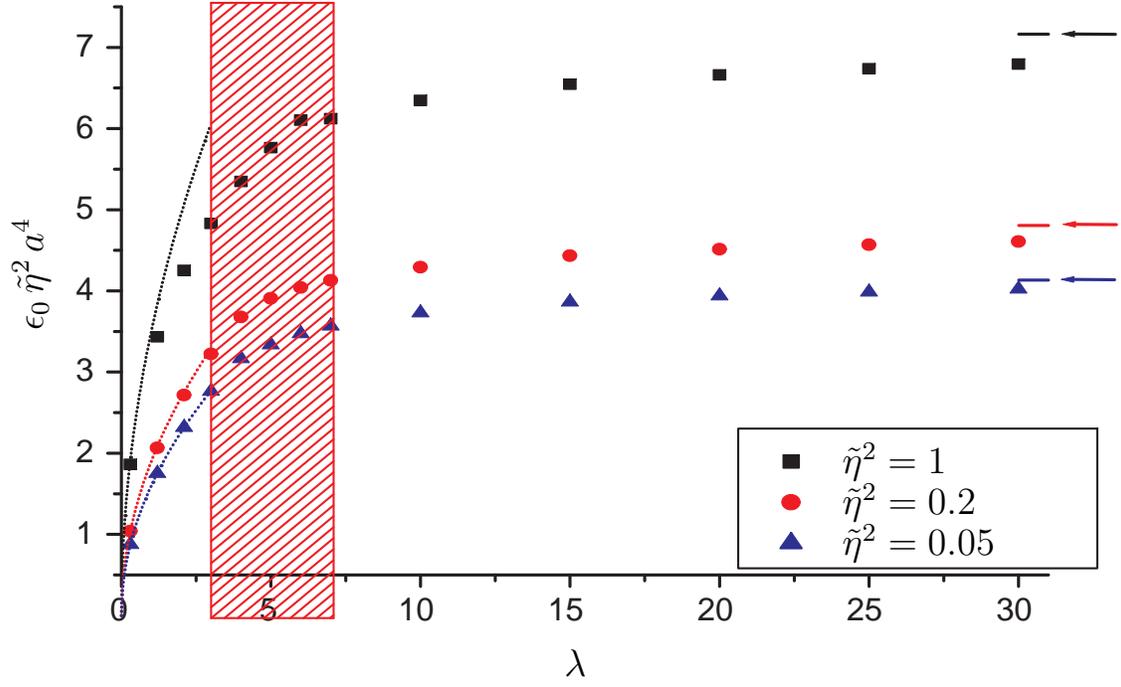}
	\caption{Optimized energy density as a function of $\lambda$ obtained from the simulation on 
	         a $16^3$ lattice for three different values of $\tilde{\eta}^2$.  
	         The red shaded area corresponds to the phase transition region for all values of $\tilde{\eta}^2$. 
	         The dotted lines show the predicted analytical strong coupling behavior. The arrows indicate
	         the expected asymptotic behavior for weak coupling which is a constant independent of $\lambda$.
	         \label{fig:E_dens_16x16x16}} 
\end{figure}

\begin{figure}[h]
	\centering
    	    \includegraphics[width=0.9\textwidth]{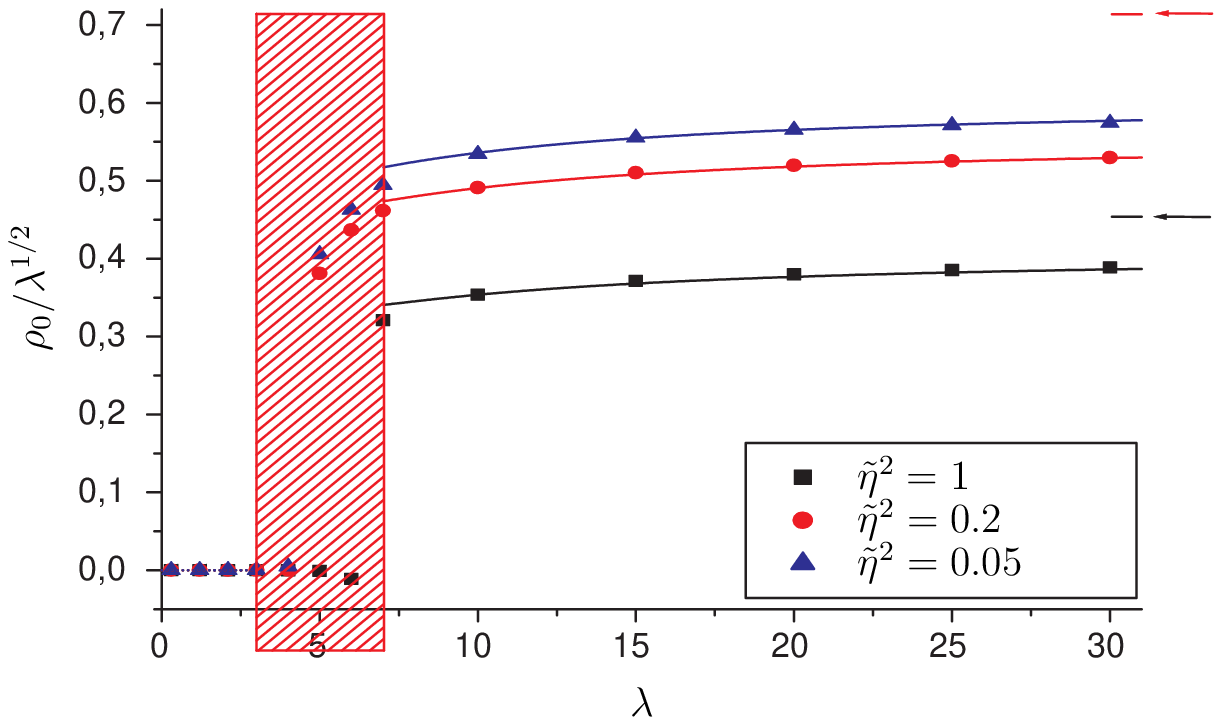}
	\caption{Optimal wave functional parameter $\rho_0(\lambda,\tilde{\eta})$ as a function of $\lambda$ 
	         obtained from the simulation on a $16^3$ lattice for three different values of $\tilde{\eta}^2$.
	         The red shaded area corresponds to the phase transition region for all values of $\tilde{\eta}^2$. 
	         The dotted lines show the predicted analytical strong coupling behavior. The arrows indicate
	         the expected asymptotic behavior for weak coupling which is proportional to $\sqrt{\lambda}$, i.e. a 
	         constant independent of $\lambda$ in the plot. 
	         The solid lines show the actual analytic parameterizations in the weak coupling regime (cf. 
	         \eqnref{eqn:fitquadraticform}).
	         \label{fig:a_16x16x16}} 
\end{figure}

\begin{figure}[h]
	\centering
	    \includegraphics[width=0.9\textwidth]{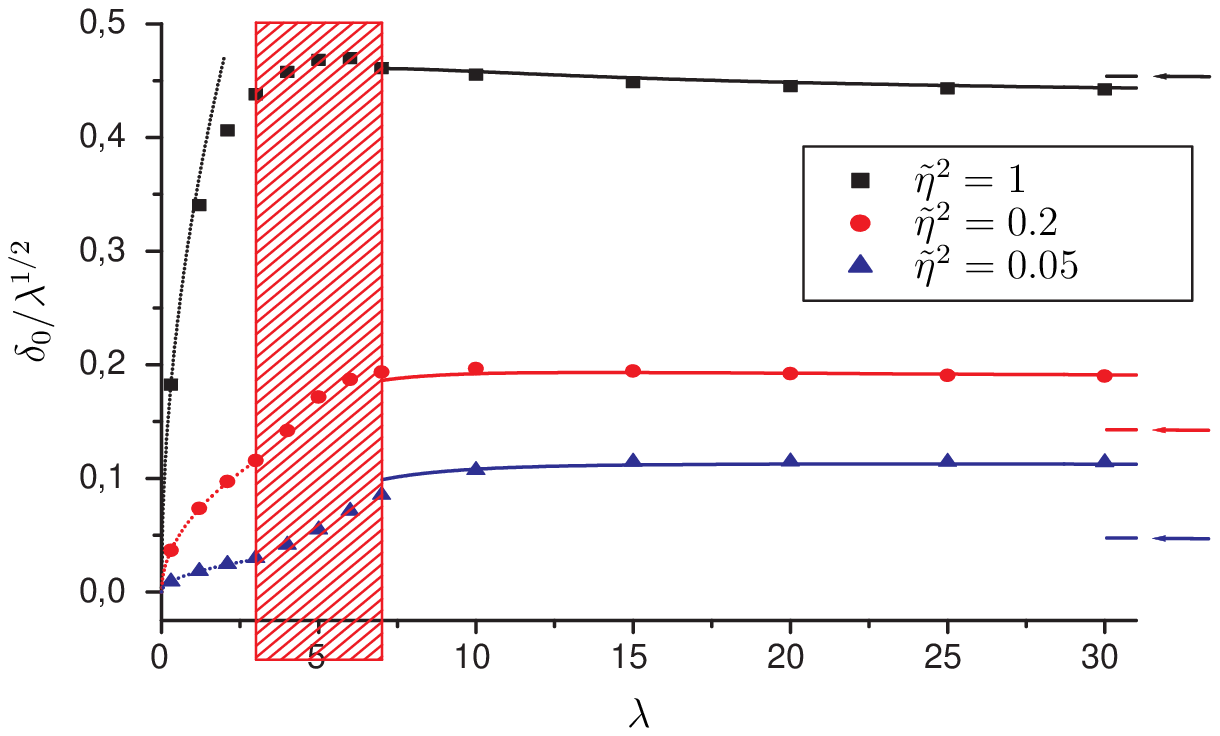}
	\caption{Optimal wave functional parameter $\delta_0(\lambda,\tilde{\eta})$ as a function of $\lambda$ 
	         obtained from the simulation on a $16^3$ lattice for three different values of $\tilde{\eta}^2$.
	         The red shaded area corresponds to the phase transition region for all values of $\tilde{\eta}^2$. 
	         The dotted lines show the predicted analytical strong coupling behavior. The arrows indicate
	         the expected asymptotic behavior for weak coupling which is proportional to $\sqrt{\lambda}$, i.e. a 
	         constant independent of $\lambda$ in the plot. 
	         The solid lines show the actual analytic parameterizations in the weak coupling regime (cf. 
	         \eqnref{eqn:fitquadraticform}).
	         \label{fig:b_16x16x16}} 
\end{figure}

Next we apply the described optimization method to the nlc
Hamiltonian. The optimized energy density is  presented in \figref{fig:E_dens_16x16x16}
as a function of $\lambda$ for different
values of $\tilde{\eta}^2$. The $1/\tilde{\eta}^2$ divergence is scaled out.
The curve has a $\sqrt{\lambda}$ behavior for strong coupling
and is independent of $\lambda$ for weak coupling as found in 
\secref{sec:AnalyticalSolutions}
\bea
\epsilon_0|_{\text{strong coupling}} &=& \frac{1}{\tilde{\eta}^2}
               \frac{1}{a_{\bot}^4} 
               \left(\frac{3}{2}+2~\tilde{\eta}^2\right)\sqrt{\lambda}
\nonumber\\
\epsilon_0|_{\text{weak coupling}} &=&\frac{1}{a_{\bot}^4}  \frac{6}{\tilde{\eta}^2} \frac{1}{N_-N_{\bot}^2}  \nn\\
& & \sum\limits_{\vec{k}} \left[\tilde{\eta}^2 \sin\left(\frac{k_1}{2}\right)^2+\tilde{\eta}^2 \sin\left(\frac{k_2}{2}\right)^2+ \sin\left(\frac{k_-}{2}\right)^2 \right]^{1/2} \;.\nn\\
& &
\eea
In Figs.~\ref{fig:a_16x16x16} and \ref{fig:b_16x16x16}, 
we present the variationally optimized wave functional
parameters $\rho_0$ and $\delta_0$ as a function of $\lambda$ for different values 
of $\tilde{\eta}^2$. The parameters are divided by a factor $\sqrt{\lambda}$
such that they become constant in the weak coupling limit ($\lambda \rightarrow \infty$).
The uncertainties on the variational parameters are typically $5\%$ and are
larger in the region where the Hamiltonian 
with the adjoint plaquette in $(-k)$-direction induces a phase
transition. Therefore, in principle only couplings
in the weak coupling region above $\lambda =7$ are meaningful
where the $Z(2)$ symmetry is spontaneously broken. 

By using the strong coupling solution from \eqnref{eqn:StrongCouplingSolution} and the diagonal 
part of the covariance matrix at $\Delta \vec{x}=\vec{0}$ of the weak coupling solution \eqnref{eqn:WeakCouplingSolution},
we get analytically the following estimates for $\rho_0$ and $\delta_0$
\bea
\rho_0(\lambda,\tilde{\eta})&=&\left\{
\begin{array}{lcr}
 0 & \mathrm{for} & \lambda<<1 \\
 \sqrt{\lambda}~\gamma_{\tilde{\eta}}(\vec{0}) \phantom{~\tilde{\eta}^2} & \mathrm{for} & \lambda>>1
\end{array}
\right.\nonumber\\
\delta_0(\lambda,\tilde{\eta})&=&\left\{
\begin{array}{lcr}
\frac{1}{3}~\lambda~\tilde{\eta}^2 & \mathrm{for} & \lambda<<1 \\
 \sqrt{\lambda}~\tilde{\eta}^2~\gamma_{\tilde{\eta}}(\vec{0}) & \mathrm{for} & \lambda>>1
\end{array}
\right.\nn\\
&\gamma_{\tilde{\eta}}(\vec{0})& \left\{
\begin{array}{lcr}
\sim 0.038/\tilde{\eta} & \mathrm{for}&\tilde{\eta} \rightarrow 0\\
=0.454 & \mathrm{for}& \tilde{\eta} \rightarrow 1
\end{array}
\right.
~.
\label{eqn:SupposedCoefficients}
\eea
The variationally determined parameters are in good agreement with the analytic
predictions in the strong coupling regime.
In the weak coupling regime the optimal parameters
differ from the analytical estimates \eqnref{eqn:SupposedCoefficients}. In both cases
the analytic predictions disagree more 
for small $\tilde{\eta}$. This is natural, since the light cone limit
$\tilde{\eta} \rightarrow 0$ builds up correlations among plaquettes separated
along the longitudinal direction. The parameters optimizing
our product of single plaquette wave functionals effectively describe these
correlations and adopt values which differ  from the weak coupling estimate 
given by the diagonal entries of the covariance matrix
\eqnref{eqn:SupposedCoefficients}. 

\begin{figure}[h]
	\centering
  \includegraphics[width=0.9\textwidth]{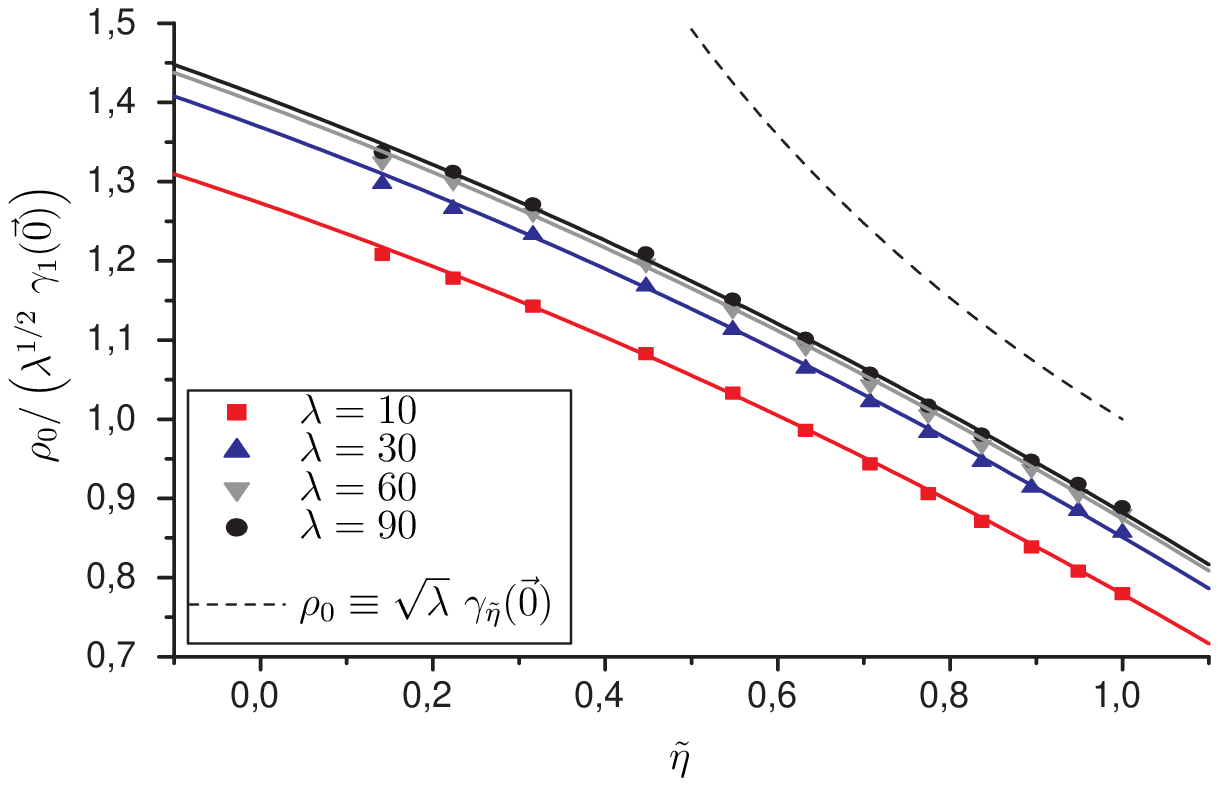}
	\caption{Optimal wave functional parameter $\rho_0(\lambda,\tilde{\eta})$ as a function of 
	         $\tilde{\eta}$ obtained from the simulation on a $16^3$ lattice for four different 
	         values of $\lambda$. The expected $\lambda^{1/2}~\gamma_{1}(\vec{0})$ behavior 
	         for the equal time Hamiltonian with $\tilde{\eta}=1$ is scaled out. The solid lines 
	         show the analytical parameterizations. The dotted line corresponds to the ``naive" 
	         analytical weak coupling prediction.
	         \label{fig:rho0ofeta_16x16x16}} 
\end{figure}

\begin{figure}[!ht]
	\centering
 	    \includegraphics[width=0.9\textwidth]{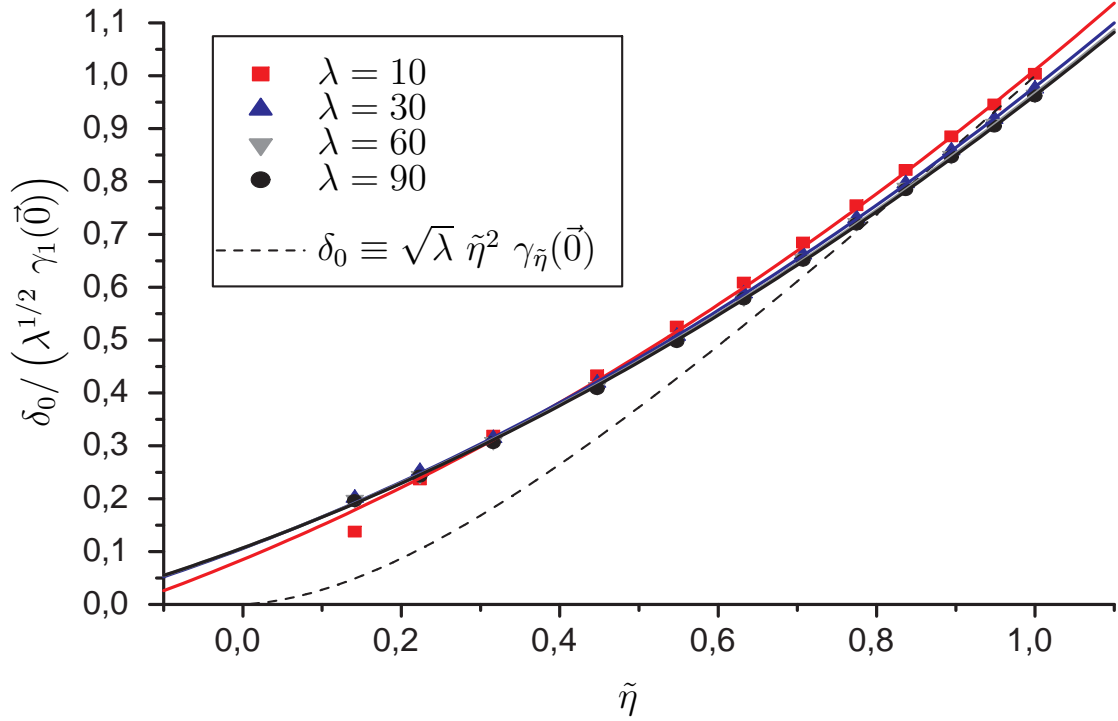}
	\caption{Optimal wave functional parameter $\delta_0(\lambda,\tilde{\eta})$ as a function of 
	         $\tilde{\eta}$ obtained from the simulation on a $16^3$ lattice for four different 
	         values of $\lambda$. The expected $\lambda^{1/2}~\gamma_{1}(\vec{0})$ behavior 
	         for the equal time Hamiltonian with $\tilde{\eta}=1$ is scaled out. The solid lines show the analytical 
	         parameterizations. The dotted line corresponds to the ``naive" analytical weak coupling prediction. 
	         \label{fig:delta0ofeta_16x16x16}} 
\end{figure}

In the following we analyze the $\tilde{\eta}$ dependence of the 
optimal wave functional parameters for fixed values of $\lambda$
which lie in the physical relevant region above $\lambda=7$. 
We show in Figs.~\ref{fig:rho0ofeta_16x16x16} and \ref{fig:delta0ofeta_16x16x16}
the optimal wave functional parameters 
$\rho_0, \delta_0$ 
divided by $\sqrt{\lambda}~\gamma_{1}(\vec{0})$, 
i.e. $\sqrt{\lambda}~\gamma_{\tilde{\eta}}(\vec{0})$ for $\tilde{\eta}=1$, which 
is the expected behavior for the equal time Hamiltonian. This way we can show the variations
of the wave functional parameters in the light cone limit.  
For a direct comparison, we plot the analytical weak coupling prediction 
\eqnref{eqn:SupposedCoefficients} by dotted lines in the same figures. 
The analytical results for $\rho_0$ (\eqnref{eqn:SupposedCoefficients}) 
overestimate the variationally determined values,
whereas the analytical predictions for $\delta_0 $ (\eqnref{eqn:SupposedCoefficients}) 
underestimate the optimized parameters as a function of $\tilde{\eta}$. Here again, 
the large difference for small $\tilde{\eta}$ originates from the effective description of
long range correlations by the parameters of our ground state wave functional in this parameter region.  
For sufficiently large values of $\lambda$,  
the  $\tilde{\eta}$ behavior for $\rho_0$ and $\delta_0$ becomes  universal and independent of $\lambda$.
We determine functions $f_\rho$ and $f_\delta$ which describe the deviations of the variationally optimized
wave functional parameters from the the weak coupling limit $\sqrt{\lambda}~\gamma_{1}(\vec{0})$
at $\tilde{\eta}=1$ (cf. Figs.~\ref{fig:rho0ofeta_16x16x16} and \ref{fig:delta0ofeta_16x16x16})
\bea
\rho_0(\lambda,\tilde{\eta}) &=& \sqrt{\lambda}~\gamma_1(\vec{0})~f_\rho(\lambda,\tilde{\eta}) \nn\\
\delta_0(\lambda,\tilde{\eta})& =& \sqrt{\lambda}~\gamma_1(\vec{0})~f_\delta(\lambda,\tilde{\eta})\;.
\eea
In the extreme weak coupling limit $\lambda \rightarrow \infty$ and
close to $\tilde{\eta} \rightarrow 1$, each of the functions $f_{\rho}$ and $f_{\delta}$
may be described by linear functions of $\tilde{\eta}$.  Therefore, it is reasonable to
assume that $f_{\rho}$ and $f_{\delta}$ can be approximated by
expansions around $\lambda \rightarrow \infty$ and $\tilde{\eta}=1$ 
\bea
f_i(\lambda,\tilde{\eta})& =& c_{0,i}\left[1+
\frac{c_{1,i}}{\lambda} + c_{2,i}~(1-\tilde{\eta})
+\frac{c_{3,i}}{\lambda^2}+ c_{4,i}~\frac{(1-\tilde{\eta})}{\lambda}
+ c_{5,i}~(1-\tilde{\eta})^2
\right]\nn\\ 
i&=& \rho,\delta
\;.
\label{eqn:fitquadraticform}
\eea
The coefficients $c_{0,i}$ represent the effective single plaquette
equal time wave functional parameters.  
A good fit of the parameters
$c_{0,i},...,c_{5,i}$ minimizing $\chi^2$ in the range $\lambda\in[10,95]$ and
$\tilde{\eta}\in[0.15,1]$ gives the coefficients tabulated in \tabref{tab:fitcoeff}.
\begin{table}[hb]
\center
\begin{tabular}{c|c|c|c|c|c|c}
i & $c_{0,i}$ & $c_{1,i}$ & $c_{2,i}$ & $c_{3,i}$ & $c_{4,i}$ & $c_{5,i}$ \\ \hline
$\rho$   & 0.90 & -1.74 &  0.72 &  4.06 & -0.40 & -0.14 \\ \hline
$\delta$ & 0.95 &  0.93 & -1.21 & -3.22 & -0.83 &  0.32 
\end{tabular}
\caption{Coefficients of \eqnref{eqn:fitquadraticform} obtained from 
least square minimization. \label{tab:fitcoeff}}
\end{table}
This analytical parameterization of the ground state wave functional 
allows to 
smoothly interpolate between ground state wave functionals belonging to 
different coupling constants and different values of $\tilde{\eta}$ in 
the physical relevant coupling constant region.
Furthermore, the given form induces generically a vanishing
expectation value of $P_{-,lat}$ which makes it optimal for the use in
a guided diffusion Monte Carlo as discussed in 
\secref{sec:EffNLCH}. Since it is an approximation to the exact ground
state it may be used for further qualitative investigations:
In a forthcoming paper we plan to determine hadronic cross
sections by simulating how a color dipole moving along the light cone
hits a neutral hadron localized at $x^-=0$. 
With the parameterization of \eqnref{eqn:fitquadraticform} we are able to extrapolate
the parameters of the wave functional to $\tilde{\eta}=0$ 
\bea
\rho_0(\lambda,0)&=& \left(0.65-\frac{0.87}{\lambda}+\frac{1.65}{\lambda^2}\right)\sqrt{\lambda} \nn\\
\delta_0(\lambda,0)&=& \left(0.05+\frac{0.04}{\lambda}-\frac{1.39}{\lambda^2}\right)\sqrt{\lambda} \;.
\eea
At $\tilde{\eta}=0$,
the color dipole can be represented by a longitudinal-transversal Wilson loop extended in $x^-$ direction 
and the simplified target can be modeled by a transverse plaquette. Varying the impact 
parameter one can sample the correlation function of the two gauge invariant objects and 
thereby obtain the profile function.


\FloatBarrier

\section{Conclusions and outlook}
\label{sec:OutlookAndConclusions}
Light cone coordinates are especially suited to parameterize high energy
reactions for which perturbative QCD calculations have reached an unprecedented
accuracy. In this paper we have addressed 
the question how to include non-perturbative features of
QCD on the light cone. We propose to use
lattice gauge theory formulated in exactly these coordinates. 

We start from the standard lattice action written in terms of
near light cone coordinates such that the continuum action is
recovered in leading order in the lattice spacing.  
The distance to
the light cone is tuned by the adjustment of an external parameter
$\eta$.  A transition to Euclidean time in this framework turns out
to be problematic from a numerical point of view due to the fact
that the Euclidean action remains complex which means that the
integrand of the path integral cannot be interpreted as a probability
measure anymore. Similar problems for QCD at finite baryonic
density are generally referred to as the sign problem for which, up to the
moment, no solution is known. In our case, this problem can be
circumvented by applying the following strategy.  We stay in
Minkowski time, but switch to a Hamiltonian formulation of lattice gauge theory.  Then
the time
evolution operator can be analytically continued to imaginary times and
acts as a projector onto the exact ground state when it is
applied to a trial state with a non-vanishing overlap with the exact ground state.
Hence, instead of sampling
the Euclidean path integral one manipulates a probability distribution
for the product of the exact ground state wave functional
and a guidance wave functional in a Quantum Diffusion Monte Carlo
algorithm. 
For an improvement of the convergence of the diffusion Monte Carlo, the guidance wave
functional should be sufficiently close to the exact ground state. The
main goal of the present paper has been to develop a convenient and numerically realizable 
ground state projection operator and to propose such a guidance wave functional.

We first work out the more obvious continuum formulation.  
The
continuum near light cone Hamiltonian has an asymmetry in the longitudinal and
transversal fields. The transversal fields are ``enhanced" in the
Hamiltonian in comparison to the longitudinal ones by a factor of
$1/\eta^2$ which is due to the underlying Lorentz transformation of the chromo
magnetic and chromo electric fields.  
Furthermore, the obtained near light cone Hamiltonian
is similar to the classical Hamiltonian of a charged particle moving
in an electromagnetic background field, which contains terms linear in
the particle momentum.  Such terms yield complex branching ratios in a 
Quantum Monte Carlo algorithm which cannot be interpreted as probabilities 
and make it fail.  However, this problem can be avoided.  Linear terms 
 in the QCD near light cone Hamilton operator can be compensated by  
the generator of longitudinal translations.  The QCD ground
state is translation invariant, i.e. an eigenstate of the longitudinal
translation operator with zero eigenvalue. Since the  Hamiltonian commutes with
the longitudinal momentum operator, the longitudinal
momentum is not affected by time evolution.  Therefore, one is able to
construct an effective Hamiltonian feasible for a Quantum Diffusion
Monte Carlo having the same ground state as the exact Hamiltonian
by adding the longitudinal momentum operator to the exact Hamiltonian.

Having checked feasibility in the continuum we derive the lattice Hamiltonian
from the action via the transfer matrix method.  We allow in general
different lattice spacings in longitudinal and transversal
directions.  It is remarkable that the parameter $\eta$
controlling the distance to the light cone multiplies
the lattice anisotropy parameter $\xi$ which represents
the ratio of the longitudinal lattice spacing and the transversal
lattice spacing. Since these two parameters always appear together,  
there is no difference between the light
cone limit and the anisotropic lattice limit. 
We can construct an effective Hamiltonian
similar to the continuum case by adding the effective longitudinal
momentum operator to the lattice Hamiltonian.

We analytically compute the lattice ground state wave functional of the
effective Hamiltonian in the strong and weak coupling limit.  In the
strong coupling limit we obtain a product of single plaquette
wave functionals similar to the equal time scenario.  In the weak
coupling limit, the solution is equivalent to the solution of the 
near light cone Hamiltonian with abelian fields, i.e. 
it is a multivariate Gaussian wave functional with a covariance matrix 
weighting correlations of field strengths at different spatial separations. 

Motivated by the strong and weak coupling solutions, 
we have constructed an effective ground state wave functional which smoothly
interpolates between these two extreme results and which can be
used as a guidance wave functional for a Quantum Diffusion Monte Carlo algorithm.  
It is a variational ansatz for the whole coupling range which contains a product 
of single plaquette wave functionals with two variational parameters.
We have variationally optimized the parameters by minimizing the energy expectation
value of the effective Hamiltonian with respect to the ground state wave functional. 
The effective ground state wave functional serves as a starting point for further qualitative
explorations. It can also be extrapolated to $\tilde{\eta}=0$ and may be used to simulate
correlation functions which appear in hadronic cross sections.

The effective ground state wave functional can be improved by allowing
also long range correlations in the wave functional. This is motivated 
by the observation that in the weak coupling limit, the covariance 
matrix elements of the analytical ground state wave functional which connect
longitudinally separated spatial points become more and more important.
An exponential ansatz then may contain plaquettes which
are connected back and forth via long strings of gauge links. 
In the light cone limit the energetically  
most favorable string configurations are elongated along the 
minus direction. Such an ansatz may interpolate in the whole
coupling range by allowing a covariance matrix with
adjustable parameters.  Numerical techniques \cite{Beccaria:2000eg}
exist for a guided random walk in parameter space. 
So it may be possible to construct on the
basis of an improved weak coupling solution a reasonable numerical
procedure to obtain a good ground state for the effective lattice
Hamiltonian.

There have been strong advances in light cone physics recently in 
string and supersymmetric theory \cite{Green:1987sp,Green:1987mn}. A careful study of near 
light cone theory in lattice QCD may supplement this successful work.

\begin{acknowledgments}
We are grateful to the Max-Planck-Institut f\"ur Kernphysik Heidelberg for providing us
with resources on the Opteron cluster. D.~G. acknowledges funding by the European Union 
project EU RII3-CT-2004-506078 and the GSI Darmstadt. E.V.~P. thanks the Russian Foundation 
RFFI for the support in this work. E.-M.~I. is supported by DFG under contract FOR 465 (Forschergruppe
 Gitter-Hadronen-Ph\"anomenologie).
\end{acknowledgments}

\appendix
\section{Some useful commutator relations}
\label{app:CommRel}
In this section we collect some useful formulae for the computation
of matrix elements. First, we want to apply an arbitrary kinetic energy
operator to the unperturbed strong coupling ground state $\ket{\Psi_0^{(0)}}$ (cf. \secref{sec:StrongCouplingSolution}) multiplied 
by an arbitrary function of the links 
\bea
\sum\limits_{j,a,\vec{y}} c_j \Pi_j^a(\vec{y})^2 f(\{U\}) \ket{\Psi_0^{(0)}} &=& 
  \sum\limits_{j,a,\vec{y}} c_j \left[ \Pi_j^a(\vec{y})^2, f(\{U\}) \right] \ket{\Psi_0^{(0)}} \nn\\
&=& 
  \sum\limits_{j,a,\vec{y}} c_j \Pi_j^a(\vec{y}) \left[ \Pi_j^a(\vec{y}), f(\{U\}) \right] \ket{\Psi_0^{(0)}} \nn\\
&=& 
  \sum\limits_{j,a,\vec{y}} c_j \left[\Pi_j^a(\vec{y}),\left[ \Pi_j^a(\vec{y}), f(\{U\}) \right]\right] \ket{\Psi_0^{(0)}} \; .
\eea
The following double commutators are of special interest
\bea
\lefteqn{
\sum\limits_{j,a,\vec{y}} c_j \left[ \Pi_j^a(\vec{y}),\left[ \Pi_j^a(\vec{y}), \Tr{\Real{ U_{kl}(\vec{x})}}\right] \right] } \nn\\
&=& \frac{3}{2} \left(c_k + c_l \right) \Tr{\Real{U_{kl}(\vec{x})}}
\label{eqn:DoubleCommTrRe}
\eea
and
\bea
\lefteqn{
\sum\limits_{j,a,\vec{y}} c_j \left[ \Pi_j^a(\vec{y}),\left[ \Pi_j^a(\vec{y}),\left( \Tr{~\Real{ U_{kl}(\vec{x})}}\right)^2\right] \right] } \nn\\
&=& 4 \left(c_k + c_l \right) \left[\left(\Tr{\Real{U_{kl}(\vec{x})}}\right)^2-1\right]
\; .
\label{eqn:DoubleCommTrRe2}
\eea
For the elementary plaquette, we have the following commutation relation
\bea
\left[\wh{\Pi}_j^a(\vec{y}),\Tr{ \Real{ U_{kl}(\vec{x}) } }\right]
&=&\phantom{-} \mathrm{i}\;\Tr{\frac{\sigma^a}{2} \; \Imag{ U_{kl}(\vec{x})} } \delta_{\vec{y},\vec{x}}\delta_{jk} \nn\\
& &+ \mathrm{i}\;\Tr{ \frac{\sigma^a}{2} \; \Imag{ U_k^\dagger(\vec{x}) U_{kl}(\vec{x}) U_k(\vec{x})}} \delta_{\vec{y},\vec{x}+\vec{e}_k}\delta_{jl} \nn\\
& &-\mathrm{i}\;\Tr{\frac{\sigma^a}{2}\;\Imag{U_l^\dagger(\vec{x})U_{kl}(\vec{x})U_l(\vec{x}) } } \delta_{\vec{y},\vec{x}+\vec{e}_l}\delta_{jk} \nn\\
& &-\mathrm{i}\;\Tr{\frac{\sigma^a}{2}\;\Imag{ U_{kl}(\vec{x}) } } \delta_{\vec{y},\vec{x}}\delta_{jl}
\; .
\label{eqn:CommElemPlaq}
\eea
In the following we assume an exponential ground state wave functional with exponent $F(\{U\})$ 
where $F(\{U\})$ is some arbitrary real valued functional of the links and $\ket{\Psi_0^{(0)}}$  is 
the unperturbed strong coupling ground state (cf. \secref{sec:StrongCouplingSolution})
\bea
\Big|\Psi_0 \Big\rangle &=& \exp\left[F(\{U\})\right] \ket{\Psi_0^{(0)}} \nn\\
\Rightarrow \Big\langle \Psi_0 \Big| &=& \bra{\Psi_0^{(0)}} \exp\left[F(\{U\})\right]~.
\label{eqn:ArbExpGroundState}
\eea 
Then, the expectation value of the color trace of the momentum operator $\Pi_j^a(\vec{y})$ squared with respect to the ground state \eqnref{eqn:ArbExpGroundState} is given by 
\bea
\lefteqn{
\sum\limits_{j,a,\vec{y}} c_j
\bra{\Psi_0}\Pi_j^a(\vec{y})^2\ket{\Psi_0} =} \nn\\
 & &\sum\limits_{j,a,\vec{y}} c_j \bra{\Psi_0}\onehalf\left[\Pi_j^a(\vec{y}),\left[\Pi_j^a(\vec{y}),F(U)\right]\right]\ket{\Psi_0}~.  
\eea
The expectation value of the momentum operator $\Pi_j^a(\vec{y})$ times an arbitrary functional $G(\{U\})$ of the links is given by
\bea
\Big\langle\Psi_0\Big| \Pi_j^a(\vec{y})~G(\{U\})\Big|\Psi_0\Big\rangle 
 &=&\phantom{-}\Big\langle \Psi_0^{(0)}\Big| \exp\Big[F(\{U\})\Big]~\Pi_j^a(\vec{y})
      ~G(\{U\})\nn\\
 & & ~~~~~~~~~~~~~~~~~~~~~~~~\cdot\exp\Big[F(\{U\})\Big] \Big| \Psi_0^{(0)}\Big\rangle \nn\\
 &=&-\Big\langle \Psi_0^{(0)}\Big|\left[\Pi_j^a(\vec{y}),\exp\Big[F(\{U\})\Big]\right]~
      ~G(\{U\})\nn\\
 & &  ~~~~~~~~~~~~~~~~~~~~~~~~\cdot\exp\Big[F(\{U\})\Big]\Big| \Psi_0^{(0)}\Big\rangle \nn\\ 
 &=&-\Big\langle \Psi_0^{(0)}\Big|\exp\Big[F(\{U\})\Big]
      ~G(\{U\})\nn\\
 & & ~~~~~~~~~~~~\cdot\left[\Pi_j^a(\vec{y}),\exp\Big[F(\{U\})\Big]\right]\Big| \Psi_0^{(0)}\Big\rangle \nn\\
 &=&-\Big\langle \Psi_0^{(0)}\Big| \exp\Big[F(\{U\})\Big]~
      G(\{U\})~\Pi_j^a(\vec{y})\nn\\
 & & ~~~~~~~~~~~~~~~~~~~~~~~~\cdot\exp\Big[F(\{U\})\Big] \Big| \Psi_0^{(0)}\Big\rangle \nn\\           
 &=&-\Big\langle\Psi_0\Big| G(\{U\})~\Pi_j^a(\vec{y})\Big|\Psi_0\Big\rangle 
\; .
\eea

\section{Derivation of the transfer matrix operator T}
\label{sec:TransferMatrix}

In this appendix we construct the transfer matrix operator $\mathbf{T}$ propagating a spatial lattice configuration from one time slice to the next
and the Hilbert space on which it acts. 
Operators are written explicitly in boldface.  
The Hilbert space on which $\mathbf{T}$ operates contains general states
$\ket{\Psi}$ which can be expanded in link states:
\bea \ket{\Psi}=\int \mathcal{D}\mathcal{ U } \;
\Psi(\mathcal{U}) \ket{\mathcal{ U }} \; .
\label{eqn:BasisStateExp}
\eea 
The measure is
$\mathcal{D}\mathcal{U}$ in \eqnref{eqn:BasisStateExp} refers to the
correspondent product of $SU(2)$ Haar measures 
\bea
\mathcal{D}\mathcal { U } &=& \prod\limits_{\vec{x},j} dU_j\left(
\vec{x} \right) \;.
\eea 
The inner product in this Hilbert space is given
by 
\bea 
\bra{\Psi^\prime}\left.\Psi\right\rangle= \int
\mathcal{D}\mathcal{U} \;
\Psi^\prime(\mathcal{U})^{*}\;\Psi(\mathcal{U}) \;.
\eea 
We define the operator $\mathbf{T}$ such that its matrix elements in the link basis 
are given by the transfer matrix \eqnref{eqn:Top1}
\bea 
\bra{\mathcal{U}(x^{\prime+})}\mathbf{T}
\Big|\mathcal{U}(x^+)\Big\rangle\equiv T(x^{\prime+},x^+)  \; .
\label{eqn:DefTransMatOp}
\eea 
The path integral for finite lattice of
$N_\tau$ time slices with periodic boundary conditions can be written
as the trace of the $N_\tau$-fold product of transfer matrices 
\bea
\int \prod\limits_{x}\prod\limits_{j=1,2,-} dU_{j}(x)e^{\mathrm{i} S_{lat}}=
\Tr{ \mathbf{T}^{N_\tau}} \; .
\eea 
The transfer-matrix operator
$\mathbf{T}$ is 
related to the Hamiltonian, the generator of time
translations  
\bea 
\mathbf{T}=
e^{-\mathrm{i} a_+ \mathbf{H}} \hspace{0.5cm} \Rightarrow
\hspace{0.5cm} \mathbf{H}= \lim_{a_+\rightarrow
0}-\frac{1}{\mathrm{i}a_+}\log\left(\mathbf{T}\right) \; .
\label{eqn:HamiltonianFromTM}
\eea
We define with the group elements $g_j(\vec{x}) \in SU(2)$ the following operators
\bea
\mathbf{U}_j(\vec{x}) \ket{\mathcal{ U }}&=&U_j(\vec{x}) \ket{\mathcal{ U }} ~~~~\forall j,\vec{x}\nn\\
\mathbf{R}\Bigl(g_j(\vec{x})\Bigr) \ket{\mathcal{ U }}&=&\ket{\mathcal{ U }^\prime }\nn\\ 
\label{eqn:OpDef} 
\ket{\mathcal{ U }^\prime }&=&
\ket{\ldots ,g_j(\vec{x}) U_j(\vec{x}),\ldots} \; .
\eea 
Here all links in $\ket{\mathcal{ U }^\prime }$ coincide with the correspondent 
links in $\ket{\mathcal{ U }}$ except for the link $U_j(\vec{x})$ which is left 
multiplied by $g_j(\vec{x})$.
The operator $\mathbf{R} ( g_j(\vec{x}))$ is similar to the translation operator in
Quantum Mechanics. It is a unitary operator and satisfies the group
representation property, i.e.  
\be
\mathbf{R}\Bigl(g_j(\vec{x})\Bigr)\mathbf{R}\Bigl(
g_j^\prime(\vec{x})\Bigr)=\mathbf{R}\Bigl(g_j(\vec{x})\cdot
g_j^\prime(\vec{x})\Bigr) \; .
\label{eqn:RepProp}
\ee 
The group elements $g_j(\vec{x})$ are parameterized by the exponential 
map which yields the Haar measure $dg_j(\vec{x})$
\bea
g_j(\vec{x})&=&e^{\mathrm{i} \gamma_j^a(\vec{x}) \sigma^a/2} ~,~\gamma_j^a(\vec{x}) \in \mathrm{reals} \nn\\
dg_j(\vec{x})&=&J\Bigl(\vec{\gamma}_j(\vec{x})\Bigr)\prod\limits_a
d\gamma_j^a(\vec{x}) \; .
\eea
The Jacobian $J$ is equal to unity in a neighborhood of $\vec{\gamma}_j(\vec{x})=\vec{0}$.
Introducing the momentum operators $\mathbf{\Pi}_j^a(\vec{x})$ canonically
conjugate to $\mathbf{U}_j(\vec{x})$
we have
\bea
\mathbf{R}\Bigl( g_j(\vec{x})\Bigr)&=&e^{-\mathrm{i}\gamma_j^a(\vec{x}) \mathbf{\Pi}_j^a(\vec{x})}\nn\\
g_j(\vec{x})&=&e^{\mathrm{i} \gamma_j^a(\vec{x}) \sigma^a/2}\nn\\
\left[ \mathbf{\Pi}_j^a(\vec{x}),\mathbf{U}_{j^\prime}(\vec{x}^\prime) \right]
&=& \phantom{-}
\frac{\sigma_a}{2}~\mathbf{U}_j(\vec{x})~\delta_{j,j^\prime}~\delta_{\vec{x},\vec{x}^\prime}
\; , \nn\\ \left[
\mathbf{\Pi}_j^a(\vec{x}),\mathbf{U}_{j^\prime}^{\dagger}(\vec{x}^\prime) \right]
&=&
-\mathbf{U}_j^{\dagger}(\vec{x})~\frac{\sigma_a}{2}~\delta_{j,j^\prime}~\delta_{\vec{x},\vec{x}^\prime}
\;.
\label{eqn:LatMomOpComm}   
\eea 
In contrast to the continuum commutation relations
\eqnref{eqn:ContComm}, the lattice momentum operators canonically
conjugate to $\mathbf{U}_j(\vec{x})$ do not commute. 
\bea
\left[ 
\left[
\mathbf{\Pi}_j^a(\vec{x}),\mathbf{\Pi}_{j^\prime}^b(\vec{x}^\prime) \right],
\mathbf{U}_k(\vec{y})\right] &=&\left[
\left[\frac{\sigma^a}{2},
~\frac{\sigma^b}{2}\right]~
,\mathbf{U}_k(\vec{y})\right]~\delta_{j,k}~\delta_{\vec{x},\vec{y}}~\delta_{j^\prime,k}~\delta_{\vec{x}^\prime,\vec{y}} \nn\\ 
&=&\mathrm{i}~\epsilon^{abc}~\left[ \frac{\sigma^c}{2}
,~\mathbf{U}_k(\vec{y})\right]~\delta_{j,j'}~\delta_{\vec{x},\vec{x}^\prime}~
\delta_{j,k}~\delta_{\vec{x},\vec{y}} \nn\\ &=&
\mathrm{i}~\epsilon^{abc}~\left[\mathbf{\Pi}_j^c(\vec{x}),~\mathbf{U}_k(\vec{y})\right]~\delta_{j,j'}~\delta_{\vec{x},\vec{x}^\prime} \; .
\eea
Since this relation is true for arbitrary $\mathbf{U}_k(\vec{y})$, we get
\bea
\left[ \mathbf{\Pi}_j^a(\vec{x}),\mathbf{\Pi}_{j^\prime}^b(\vec{x}^\prime) \right] 
&=&
\mathrm{i}~\varepsilon^{abc}~\mathbf{\Pi}_j^c(\vec{x})~\delta_{j,j^\prime}~\delta_{\vec{x},\vec{x}^\prime}
\; ,\\ \left[ \mathbf{\Pi}_j(\vec{x})^2,\mathbf{\Pi}_{j^\prime}^b(\vec{x}^\prime)
\right] &=& 0 \; .
\eea 
Note
that we have defined the translation operator on the group manifold
$\mathbf{R}$ with an opposite sign inside of the exponential in comparison with
\cite{creutz}. Our definition yields the same commutation relations as
\cite{Kogut:1974ag} which reproduce the continuum commutation
relations \eqnref{eqn:ContComm} with the gauge field $A_j^a(\vec{x})$
in the naive continuum limit. For simplicity we abandon to write quantum mechanical operators
 explicitly in boldface in the following. By using the group translation operators $R$ 
 we may write for the transfer matrix operator   
\bea 
{T} &=&\Bigg[\hspace{-0.14cm}\Bigg[\prod\limits_{\vec{x}}\int
dg_-(\vec{x})~{R}\Bigl( g_-(\vec{x})\Bigr)
~\exp\left\{\mathrm{i}\frac{2}{g^2}~\frac{a_{\bot}^2}{a_+a_-}~\Tr{
\unitop - \Real{g_-(\vec{x}}}\right\}\Bigg]\hspace{-0.14cm}\Bigg]
\times \nn \\ & & \Bigg[\hspace{-0.14cm}\Bigg[
\prod\limits_{\vec{x},k} \int dg_k(\vec{x})~{R}\Bigl(
g_k(\vec{x})\Bigr) ~\exp\left\{
\mathrm{i}~\frac{2}{g^2}~\eta^2~\frac{a_-}{a_+}~\Tr{ \unitop -
\Real{g_k(\vec{x})}}\right\} \Bigg. \Bigg.  \nn \\ & & \Bigg. \Bigg.
\hspace{1.0cm} \times ~\exp\left\{ \mathrm{i}~\frac{2}{g^2}~\Tr{
\Imag{g_k(\vec{x})}~\Imag{{U}_{-k}(\vec{x})}} \right\}
\Bigg]\hspace{-0.14cm}\Bigg] \times
\label{eqn:Top2}    \\
& & \Bigg[\hspace{-0.14cm}\Bigg[\prod\limits_{\vec{x}}
\exp\left\{-\mathrm{i}~\frac{2}{g^2}~\frac{a_+a_-}{a_{\bot}^2}~\Tr{
\unitop - \Real{{U}_{12}(\vec{x})}}\right\}
\Bigg]\hspace{-0.14cm}\Bigg] \; .  \nn
\eea 
It still has the right matrix elements \eqnref{eqn:DefTransMatOp}.
In order to arrive at \eqnref{eqn:Top2}  
one uses the fact that $R(g_j(\vec{x}))$ parameterizes the translation
in group space from $U_j(\vec{x},x^+)\rightarrow
U_j(\vec{x},x^{\prime+})$ 
\bea 
g_j(\vec{x})=
U_j(\vec{x},x^{\prime+})U_j^\dagger(\vec{x},x^+) \; .  
\eea 
Now, one
may perform the group integrations in \eqnref{eqn:Top2} explicitly.
In the limit $a_+ \rightarrow 0$, the time evolution along one
temporal step $a_+$ induces rotations $g_j(\vec{x})$ which are of the
order $a_+$ and are close to $\unitop$. This implies that the
parameters $\gamma_j^a(\vec{x})$ parameterizing these shifts are of
the order $a_+$ as well. Therefore, it is convenient to make an
expansion around $\gamma_j^a(\vec{x})=0$ up to order
$\mathcal{O}(a_+^2)$. In this limit, the Jacobian is approximately
equal to $1$ and the integrals become Gaussian integrals which can be
analytically computed. One obtains
\bea 
{H}_{\mathrm{lat}}&=& \lim_{a_+\rightarrow 0}
\left[-\frac{1}{\mathrm{i}a_+}\log\left(T\right)\right] \nn\\
&=&\sum\limits_{\vec{x}} \Bigg[\hspace{-0.14cm}\Bigg[
~\frac{g^2}{2}~\frac{1}{a_-}~\sum\limits_{k,a}~\frac{1}{\eta^2}~
\left\{~{\Pi}_k^a(\vec{x})-\frac{2}{g^2}~\Tr{\frac{\sigma_a}{2}~\Imag{{U}_{-k}(\vec{x})}}
\right\}^2 \Bigg. \Bigg. \nn\\ & & \hspace{0.5cm} \Bigg. \Bigg.
+~\frac{g^2}{2}~\frac{a_-}{a_{\bot}^2}~\sum\limits_{a}{\Pi}_-^a(\vec{x})^2
+~\frac{2}{g^2}~\frac{a_-}{a_{\bot}^2}~\Tr{\unitop-\Real{{U}_{12}(\vec{x})}}
\Bigg]\hspace{-0.14cm}\Bigg] \; . \nn\\ & & 
\eea


\end{document}